\documentclass[10pt,journal,compsoc]{IEEEtran}
\usepackage{balance}
\usepackage{cite, graphicx,amssymb,color, kbordermatrix, pict2e, multirow, rotating}
\usepackage{graphicx,amssymb}
\usepackage[tbtags]{amsmath}
\usepackage{times,amsmath}
\usepackage{subfig}
\usepackage{flushend}
\usepackage{amsmath}
\usepackage{epsfig}
\usepackage{amssymb}
\usepackage{hyperref}
\usepackage{color}
\usepackage{psfrag}
\usepackage{subfig}
\usepackage[usenames,dvipsnames]{xcolor}
\usepackage{textcomp}
\usepackage{array}
\usepackage{tikz}
\usetikzlibrary{matrix,decorations.pathreplacing}
\usepackage{tabularx,ragged2e}
\newcolumntype{C}{>{\Centering\arraybackslash}X} 
\usepackage{balance}

\usepackage{algorithmic}
\usepackage{algorithm}

\newtheorem{definition}{Definition}

\newtheorem{lemma}{Lemma}

\newtheorem{theorem}{Theorem}

\begin{document}

\title{The Design of Dynamic Probabilistic Caching with Time-Varying Content Popularity} 

\author{\IEEEauthorblockN{Jie~Gao, \IEEEmembership{Member, IEEE},   
		Shan~Zhang, \IEEEmembership{Member, IEEE},
		Lian~Zhao, \IEEEmembership{Senior Member, IEEE}, 
		and Xuemin~(Sherman)~Shen, \IEEEmembership{Fellow, IEEE} 
	}

	\IEEEcompsocitemizethanks{
		
		\IEEEcompsocthanksitem J. Gao and X. Shen are with the Department of Electrical and Computer Engineering, University of Waterloo, Waterloo, ON, N2L 3G1, Canada (e-mail: \{jie.gao, sshen\}@uwaterloo.ca). 
		
		\IEEEcompsocthanksitem L. Zhao is with the Department of Electrical, Computer, and Biomedical Engineering, Ryerson University, Toronto, ON, M5B 2K3, Canada (e-mail: l5zhao@ryerson.ca).
		
		\IEEEcompsocthanksitem S.~Zhang is with the School of Computer Science and Engineering, Beihang Univesity, Beijing 100191, China (e-mail: zhangshan18@buaa.edu.cn). She is also with the State Key Laboratory of Software Development Environment, and the Beijing Key Laboratory of Computer Networks.
	
	 	\IEEEcompsocthanksitem  This work was supported in part by the Natural Sciences and Engineering Research Council of Canada under Grant STPGP-493787 and in part by the Nature Science Foundation of China under Grant 61801011.
	}
		
}

\IEEEtitleabstractindextext{
\begin{abstract}

In this paper, we design dynamic probabilistic caching for the scenario when the instantaneous content popularity may vary with time while it is possible to predict the average content popularity over a time window. Based on the average content popularity,  optimal content caching probabilities can be found, e.g., from solving optimization problems, and existing results in the literature can implement the optimal caching probabilities via static content placement. The objective of this work is to design dynamic probabilistic caching that: i) converge (in distribution) to the optimal content caching probabilities under time-invariant content popularity, and ii) adapt to the time-varying instantaneous content popularity under time-varying content popularity.  Achieving the above objective requires a novel design of dynamic content replacement because static caching cannot adapt to varying content popularity while classic dynamic replacement policies, such as LRU, cannot converge to target caching probabilities (as they do not exploit any content popularity information). We model the design of dynamic probabilistic replacement policy as the problem of finding the state transition probability matrix of a Markov chain and propose a method to generate and refine the transition probability matrix. Extensive numerical results are provided to validate the effectiveness of the proposed design.

\end{abstract}

}

\maketitle

\IEEEdisplaynontitleabstractindextext

\IEEEpeerreviewmaketitle

\IEEEraisesectionheading{\section{Introduction}\label{s:intro}} 

\IEEEPARstart{D}{ata} traffic volume in cellular networks has experienced a tremendous growth since the deployment of the LTE, and a major portion of the traffic is related to content delivery \cite{M_XWang2014}. This feature is expected to be even more prominent in the next generation cellular network \cite{J_IParvez2018}. In this background, mobile edge caching (MEC) has attracted increasing research attention \cite{J_EMarkakis2017}-\cite{C_SZhang2018}. By placing selected contents in the cache, a base station (BS) can resolve a part of content requests locally without fetching the requested content over the backhaul. This benefits both the network and the users by alleviating the pressure on the backhaul and reducing the end-to-end delay in content delivery, respectively.

In practice, caches deployed in a network have limited sizes and can only accommodate selected contents. The problem of selecting the contents to be stored at available caches is referred to as \textit{content placement}. In the case of static caching, the cached contents will not change once the content placement problem is solved. By contrast, the cached content can be updated, e.g., as content request and download status changes, in the case of dynamic caching. The problem of selecting new content to cache while replacing existing contents is referred to as \textit{content replacement}. The term caching can include content placement, content replacement, or both. Various principles and approaches for caching can be found in the literature \cite{J_KZhang2018}-\cite{J_LQiu2019}.

When the content popularity is unknown, dynamic caching is usually adopted to allow adjustments to the content placement based on the content request and download status. With heuristic principles such as the least-frequently-used (LFU) or least-recently-used (LRU), a less popular content is replaced when a new content is accepted in classic dynamic caching. Probabilistic dynamic caching, which originates from computer networks, implements dynamic caching  probabilistically. The idea is that a user-requested content is cached with a certain probability \cite{J_STarnoi2014},\cite{C_WBao2018}. Because of the dynamic cache-and-replace process, dynamic probabilistic caching may adapt to time-varying content popularity without knowing the popularity. Moreover, dynamic probabilistic caching can achieve fair and efficient content placement in a network with low redundancy in a distributed setting \cite{J_STarnoi2014},\cite{J_IPsaras2014}. However, it is difficult to establish a bound on the performance of probabilistic caching, which is typically evaluated numerically.

When the content popularity is known, the optimal content placement can be calculated, and static caching based on the optimal placement is usually adopted. A common approach for content placement in such case is for caches at different locations in a network to cooperate and find a joint optimal caching solution. For example, Applegate~\textit{et al.} studied the placement of video-on-demand contents considering an arbitrary network topology and time-varying content demand \cite{J_DApplegate2016}. Using relaxation and approximation methods \cite{J_IDBaev2008}, the authors found a near-optimal solution that can serve all requests with significantly less bandwidth consumption when compared to LRU/LFU.  Common objectives for joint optimal caching include maximizing the cache hit probability \cite{J_NDeng2018}, maximizing the caching capacity of a network \cite{J_SZhang2017}, and minimizing the content provisioning delay \cite{SZhang2018},\cite{J_ZSU2018} or cost \cite{J_QLi2017}. Alternatively, the problem of content placement can also be formulated as a non-cooperative game \cite{J_YCui2018} or an auction \cite{J_MMangili2017} between cache servers, content providers, and/or network operators, and the solution can be found from the resulting equilibrium or outcome in a decentralized manner \cite{J_YYang2018}, \cite{C_WWang2018}.  Probabilistic content placement was adopted in many recent works in the literature of optimization-based MEC with known content popularity \cite{J_YChen2017}-\cite{J_JWen2017}. Chen~\textit{et al.} analyzed the performance of probabilistic content placement at small base stations (SBSs) and found the optimal joint caching probability to maximize the successful download probability \cite{J_YChen2017}. Li~\textit{et al.} studied the joint optimization of caching probabilities for maximizing the successful delivery probability in the scenario of an $N$-tier heterogeneous network \cite{J_KLi2018}. In \cite{JYZhou2017}, SBSs cooperate and form clusters, and an efficient solution of content placement to reduce the cooperative strategy was proposed. Liu~and~Yang studied the problem of optimal probabilistic content placement policy to maximize the area spectral efficiency in a heterogeneous network and analyzed the impact of transmission power, node density, and rate requirements on the result\cite{J_DLiu2017}. The optimal tier-level content placement with probabilistic content placement for maximizing the cache hit probability, in which BSs at the same tier are assigned the same caching probabilities, was derived in \cite{J_JWen2017}. However, most of these works do not consider dynamic content replacement as it becomes unnecessary when the exact content popularity is known. 

It is well known that content popularity can be subject to variations over time \cite{M_MZeng2018}, \cite{C_TTanaka2016}. For example, Traverso~\textit{et al.} proposed a shot noise model to capture the temporal locality in content popularity in \cite{J_STraverso2013}, while Leonardi and Torrisi analyzed LRU under the shot noise model \cite{C_ELeonardiq2015}. Garetto~\textit{et al.} proposed a unified approach to analyze caching policies, taking the temporal locality into account~\cite{J_MGaretto2016ACM}.  Given potential variations in content popularity over time, assuming the knowledge of instantaneous content popularity might not always be practical. However, the average content popularity over a time window may be obtained, e.g., from prediction \cite{C_GGursun2011}.~\footnote{Without the average content request probability information, the cache does not have many options in terms of caching strategy besides using classic policies such as LRU.} In such case, the knowledge of the average content popularity can be used to derive the target optimal overall content caching probabilities. 
It is not difficult to implement the target caching probabilities using static caching. Alternatively, one could use dynamic caching such as LRU. The former option maximizes the average cache hit ratio (among all static caching) but cannot adapt to the time-varying content popularity. The latter, by contrast, may cope with variations in the instantaneous content popularity but generally does not exploit the average content popularity information. Depending on the content request statistics, either static caching based on the target caching probabilities or dynamic caching can be the better option. The objective of this paper is to design dynamic probabilistic caching that inherit advantages from both options. Specifically, we aim at such dynamic probabilistic caching that can converge (in distribution) to target optimal content caching probabilities if the content popularity is time-invariant and can adapt to the time-varying instantaneous content popularity otherwise. To do that, we exploit the average content popularity information while designing the content replacement policy. The contributions of this work are as follows.


First, we propose the idea of dynamic probabilistic caching based on average content popularity and demonstrate that it can outperform both the static caching and classic dynamic caching such as LRU. The proposed dynamic probabilistic caching can adapt to the time-varying instantaneous content popularity and thereby increase the cache hit ratio when compared to the optimal static caching in the case of time-varying instantaneous content popularity. In addition, unlike classic dynamic caching, the proposed design can converge (in distribution) to a target set of optimal content caching probabilities in the case of time-invariant content popularity.

Second, we establish a connection between a set of caching probabilities and a probabilistic content placement policy. We show that the probabilistic content placement policy that can implement a given set of caching probabilities can be non-unique in most cases and derive a general-form solution of probabilistic content placement policies, which includes existing solutions, e.g., the one in \cite{J_BBlaszczyszyn2015}, as special cases. Moreover, we show that different probabilistic content placement policies are not equivalent when implemented in dynamic caching.

Third, we formulate the problem of designing a probabilistic content replacement policy based on average content popularity into an equivalent problem of designing the state transition probability matrix of a Markov chain and propose a method to solve the latter problem. In classic dynamic caching such as LRU, the content request statistics determine the state transitions. However, our design builds the state transition probability matrix by exploiting the average content popularity information. The proposed method finds the state transition probability matrix so that the underlying Markov chain is irreducible and ergodic, and its unique steady state implements the target probabilistic content placement policy if the instantaneous content popularity converges to the average content popularity.  

\section{System Model}

\begin{figure}[tt]
	\centering {\includegraphics[width=0.48\textwidth]{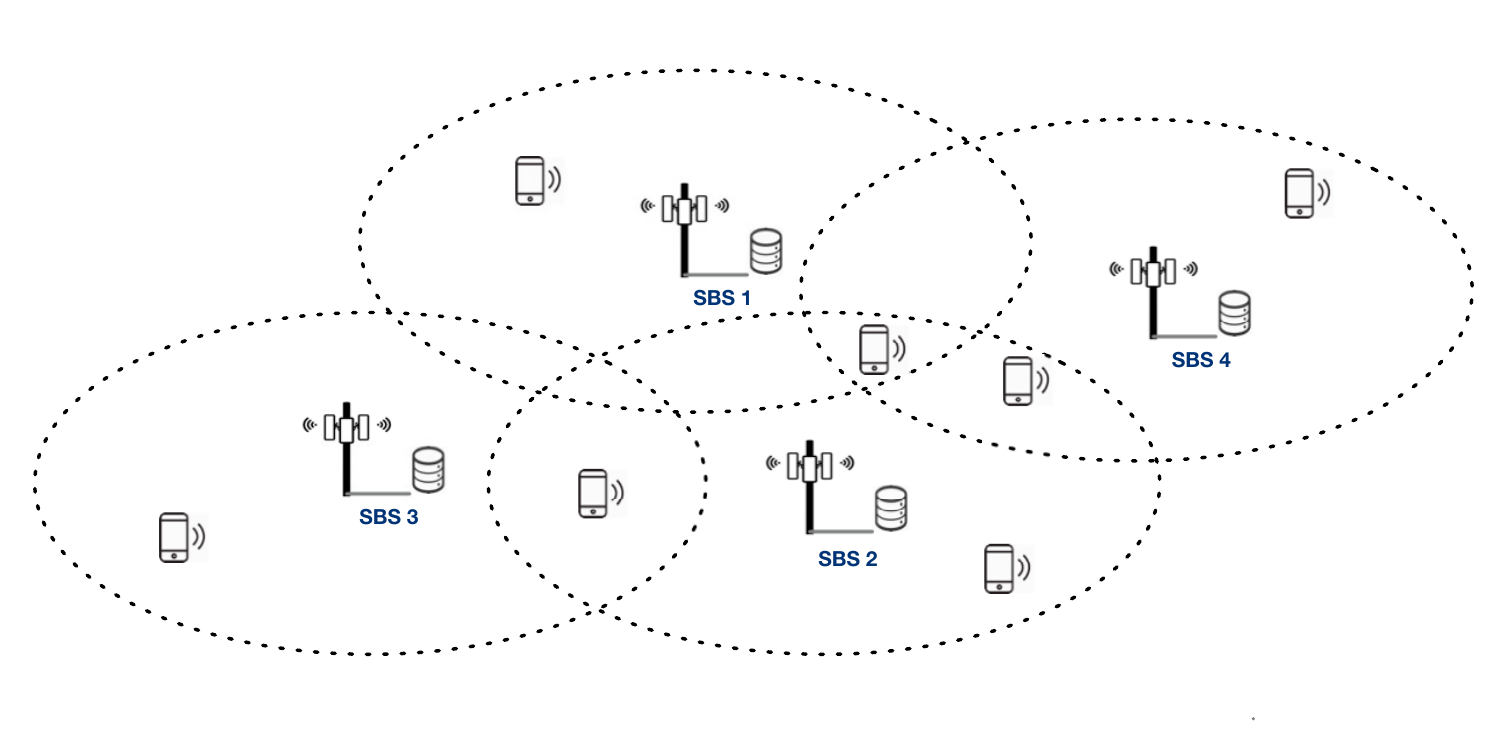}}
	\hspace{-2mm}
	\caption{An illustration of an edge caching system.}\label{f:sys}
\end{figure}
\hspace{-12mm} 
%

Consider an edge caching system (Fig.~\ref{f:sys} shows an example), where each SBS has a cache. Similar scenarios of MEC can be found in the literature, e.g., \cite{J_NGolrezaei2013}-\cite{J_JGao2018PartII}. Here we focus on the dynamic caching strategy of each individual cache. Consider a target cache with size $c$ and assume there are $N_\mathrm{f}$ contents in total. Contents have an identical length (normalized to 1), and the set of contents is denoted by $\mathcal{F}$. 
The instantaneous content request probability at the target cache can be varying with time and hard to track. However, we assume that it is possible to predict the average probability of the cache receiving a request for content $k \in \mathcal{F}$ over a predefined time window and denote this probability as $\varphi_k$~\footnote{Note that, while the average content popularity may or may not be available a priori in general, we assume that it can be predicted in our considered scenario.}. We do not consider the issue of user mobility since the impact of user mobility is reflected through the average content popularity at the target cache while our focus is not on the popularity prediction. For research on mobility management in MEC, interested readers can check references \cite{J_AGaddah2010}-\cite{C_FZhang2015}. 

As probabilistic caching is considered, contents can be cached with probabilities. Denote the probability that content $k$ is stored at the target cache as $p_k$. The cache size limit requires that 
\begin{align}
\sum\limits_{\forall k \in \mathcal{F}} p_k \leq c.
\end{align}
To maximize its cache hit ratio, an SBS would use all available cache spaces. In the sequel, we assume that all cache space is used (which implies that using cache space does not incur a cost). Depending on the specific scenario and objective, the caches in the system may find their optimal caching probabilities either locally or jointly through an optimization problem or a game using the knowledge of $\{\varphi_k\}_{k \in \mathcal{F}}$. Denote the optimal caching probabilities of the target cache by $\{p_k^\star\}_{\forall k}$.

In our model, we focus on the caching strategy after $\{p_k^\star\}_{\forall k}$ are obtained instead of finding $\{p_k^\star\}_{\forall k}$. After knowing $\{p_k^\star\}_{\forall k}$, the target cache could implement $\{p_k^\star\}_{\forall k}$ using static caching. Alternatively, the target cache could also use dynamic caching such as LRU. The former maximizes the average cache hit ratio among all static caching in the predefined time window, while the latter may cope with time-varying content popularity. We aim to design dynamic caching that can implement $\{p_k^\star\}_{\forall k}$ when the content popularity is time-invariant and can adapt when the content popularity is time-varying.

\begin{table}[t!]
	\begin{center}
		\caption{List of Symbols}\label{t:Notation}
		{\setlength{\extrarowheight}{1.5pt}
			\begin{tabularx}{0.49\textwidth}{c|X}\hline\hline
				$c$ & Cache size of the target SBS \\ \hline
				$\mathcal{F}$ &  Set of all contents \\ \hline
				$N_\mathrm{f}$ &  Number of all contents  \\ \hline
				$\varphi_k$ &  Probability that content $k$ is requested \\ \hline	
				$\mathbf{s}^{m}$ & The $m$th cache state, represented by a $N_\mathrm{f}\times 1$ vector  \\ \hline
				$\mathbf{S}$ &  The cache state matrix  \\ \hline
				$\eta^l$	& The probability of the cache in state $\mathbf{s}^l$  \\ \hline
				$\boldsymbol{\eta}$	& The state caching probability vector   \\ \hline
				$\boldsymbol{\eta}^\star$	& The target (optimal) state caching probability vector   \\ \hline
				$\mathcal{G}^m$ & The set of contents cached by cache state $\mathbf{s}^m$ \\ \hline
				$\mathcal{H}_{m}$ & The neighbor states of state $m$  \\ \hline
				$\mathcal{H}_{m}^k$ & The neighbor states of state $m$ linked by content $k$ \\ \hline
				$p_k$	& The probability of caching content $k$ \\ \hline
				$\mathbf{p}$ &  Content caching probability vector, i.e., $[p_1^\star, \dots,  p_{N_\mathrm{f}}]$  \\ \hline
				$\mathbf{p}^\star$ &  The target (optimal) content caching probability vector \\ \hline
				$\tau_{m^\prime, m}$ &  The conditional probability that the content in $\mathcal{G}^{m^\prime} \!\!\!\!-\! \mathcal{G}^m$ replaces that in $\mathcal{G}^m \!\!\!-\! \mathcal{G}^{\!m^\prime}$ \\ \hline	
				$\mathsf{S}_l$ &  The $l$th ordered sequence of states \\ \hline
				$B(l)$ &  The branch point of sequence $l$  \\ \hline
				$M(l)$ &  The merge point of sequence $l$  \\ \hline
				\hline\hline
			\end{tabularx}
		}
	\end{center}
\end{table}

%

Given the above objective, the next two sections formulate static caching in terms of probabilistic content placement policy and dynamic caching in terms of probabilistic content replacement, respectively.

\section{Probabilistic Content Placement}\label{s:ProbContentPlace} 

A cache accommodates a combination of contents, subject to its size limit, at any given time. Therefore, in probabilistic caching, it is necessary for a cache to map a given set of content caching probabilities into specific combinations of contents and the corresponding probabilities of caching these combinations. In this section, we first define cache state and then investigate static probabilistic caching by studying the connection between content caching probabilities and state caching probabilities.

\subsection{Cache states}


A cache state can be represented by a 0-1 vector: the $k$th element is 1 if content $k$ is cached and 0 otherwise. Given the cache limit $c$, the cache has $n = \binom{N_\mathrm{f}}{c}$ different states that cache exactly $c$ files. Denote the $m$th state of the cache and the set of contents cached by the $m$th state by $\mathbf{s}^m$ and  $\mathcal{G}^m$, respectively. It follows that  $\mathbf{s}^m = [s_1^m, \dots, s_{N_\mathrm{f}}^m]^\mathrm{T}$ is a $N_\mathrm{f}\times 1$ vector, where the superscript $(\cdot)^\mathrm{T}$ denotes the transpose. In the vector $\mathbf{s}^m$, element $s^m_k$ is 1 if $k \in \mathcal{G}^m$ and 0 if $k \notin \mathcal{G}^m$. We can organize all state vectors into an $N_\mathrm{f} \times n$ cache state matrix $\mathbf{S}$, which is defined as
$\mathbf{S} = [\mathbf{s}^1, \cdots, \mathbf{s}^{n}]$.

  


\subsection{State caching probability}

State caching probabilities are the probabilities assigned to the cache states. Denote the probability of cache state $\mathbf{s}^{l}$ as $\eta^l$. It follows that $\sum\limits_{l=1}^n \eta^l = 1$. 
The state caching probabilities $\{\eta^l\}_{\forall l}$ and the content caching probabilities $\{p_k\}_{\forall k}$ must satisfy the following conditions
\begin{subequations}
\begin{align}\label{e:VariableReduce}
\eta^l = 0, \;\;\forall l|\mathbf{s}^{l}_k = 0,\;\; \text{if}\;\; p_k = 1,  \\
\eta^l = 0, \;\;\forall l|\mathbf{s}^{l}_k = 1,\;\; \text{if}\;\; p_k = 0.
\end{align}
\end{subequations}
It can be seen that the number of cache states with nonzero state caching probabilities reduces for each content $k$ such that $p_k = 1$ or $p_k = 0$. The state caching probabilities can be organized into an $n \times 1$ state caching probability vector $\boldsymbol{\eta}$, as follows:
\begin{align}
\boldsymbol{\eta} \!=\! [\eta^1, \scalebox{.85}{$\dots$}, \eta^{n}]^\mathrm{T}. 
\end{align}

\subsection{Static probabilistic caching}\label{ss:ProbCaching}

Implementing a target set of content caching probabilities, e.g., $\{p_k^\star\}_{\forall k}$ is equivalent to determining the vector $\boldsymbol{\eta}$. This is because  $\boldsymbol{\eta}$ specifies the cache states and the corresponding state caching probabilities. Define the content caching probability vector $\mathbf{p} = [p_1, \dots, p_{N_\mathrm{f}}]$. The content caching probability vector $\mathbf{p}$, state caching probability vector $\boldsymbol{\eta}$, and the cache state matrix  $\mathbf{S}$ are connected through the following equation:
\begin{align}\label{e:PCacheMixRelation}
\mathbf{S}  \boldsymbol{\eta} = \mathbf{p}
\end{align}
where $\mathbf{S}$, $\boldsymbol{\eta}$,  $\mathbf{p}$ are of size $ N_\mathrm{f} \times n$,  $n \times 1$, and $N_\mathrm{f} \times 1$, respectively.

\begin{lemma}\label{l:one2multi}
	A given set of caching probabilities $\{p_k\}_{\forall k}$ could be implemented by more than one state caching probability vector $\boldsymbol{\eta}$. 
\end{lemma}

The proof of Lemma~\ref{l:one2multi} is straightforward. Since $n = \binom{N_\mathrm{f}}{c} \geq N_\mathrm{f}$, it can be seen that eq.~\eqref{e:PCacheMixRelation} is in general an under-determined system that may have more than one solution of $\boldsymbol{\eta}$, corresponding to different probabilistic content placement policies. 

Note that eq.~\eqref{e:PCacheMixRelation} incorporates the requirement that $\sum\limits_l \eta^l = 1$. This can be seen by multiplying an all-one vector of size $1\times N_\mathrm{f}$ from the left to both sides of eq.~\eqref{e:PCacheMixRelation}.

To implement a target set of content caching probabilities $\{p_k^\star\}_{\forall k}$ (equivalently, the content caching probability vector $\mathbf{p}^\star$) using static caching, the method is to randomly draw a cache state based on $\boldsymbol{\eta}^\star$. Once the cache state is drawn, there will be no change of cache state (i.e., no content replacement). As long as $\mathbf{S}  \boldsymbol{\eta}^\star = \mathbf{p}^\star$, the random draw of the cache state implements the target content caching probabilities. Lemma~\ref{l:one2multi} suggest that, the static probabilistic caching that can implement a target set of content caching probabilities is non-unique.


\begin{lemma}\label{l:SProp}
The cache state matrix $\mathbf{S}$ has the following two properties: 
\begin{itemize}
	\item $\mathbf{S}$ has full row rank;
	\item the minimum singular value of $\mathbf{S}$ is no less than $\sqrt{c}$.
\end{itemize}
\end{lemma}

\textit{Proof}: See Appendix~A.   


Lemma~\ref{l:SProp} suggests that the matrix $\mathbf{S}\mathbf{S}^\mathrm{T}$ has full rank. Therefore, given a set of optimal caching probabilities $\mathbf{p}^\star$, the solution to eq.~\eqref{e:PCacheMixRelation} can be written as:
\begin{align} \label{e:etaSolu}
\boldsymbol{\eta}^\star = \mathbf{S}^\mathrm{T} \left(\mathbf{S} \mathbf{S}^\mathrm{T} \right)^{-1} \mathbf{p}^\star + \mathbf{v},  
\end{align}
where $\mathbf{v}$ can be any vector in the null space of $\mathbf{S}$ that renders $\boldsymbol{\eta}^\star$ a valid probability vector, i.e., $\mathbf{0} \preceq \boldsymbol{\eta}^\star \preceq \mathbf{1}$ and $\mathbf{1}^\mathrm{T}\boldsymbol{\eta}^\star  = 1$. Here, the inequality sign $\preceq$ represents element-wise comparison, i.e., $\mathbf{x} \preceq \mathbf{y}$ if $x_i \leq y_i, \forall i$ (in which $x_i$ represents the $i$th element of vector $\mathbf{x}$). The solution in eq.~\eqref{e:etaSolu} includes all possible static probabilistic caching that can implement $\mathbf{p}^\star$.

\begin{figure}[!t]
	\centering \subfloat[Choosing a cache state probabilistically based on the method proposed in \cite{J_BBlaszczyszyn2015}. The result shown corresponds to one solution of $\boldsymbol{\eta}$ in eq.~\eqref{e:etaSolu}.]
	{\includegraphics[angle=0,width=0.32\textwidth]{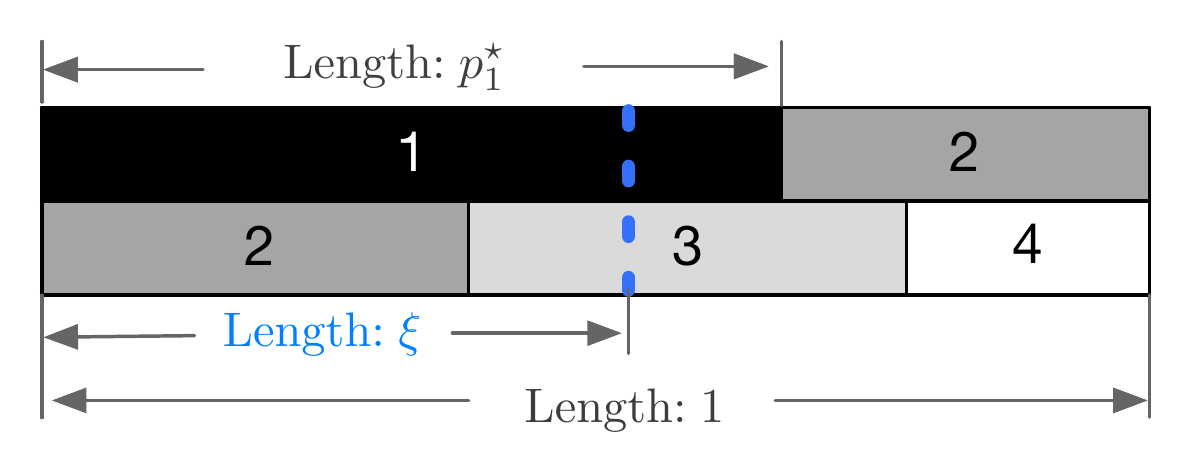}
		\label{f:HeurisSolu1}}
	\hspace{1mm} 
	\subfloat[An alternative result, which correpsonds to a different solution of $\boldsymbol{\eta}$ in eq.~\eqref{e:etaSolu}.] 
	{\includegraphics[angle=0,width=0.32\textwidth]{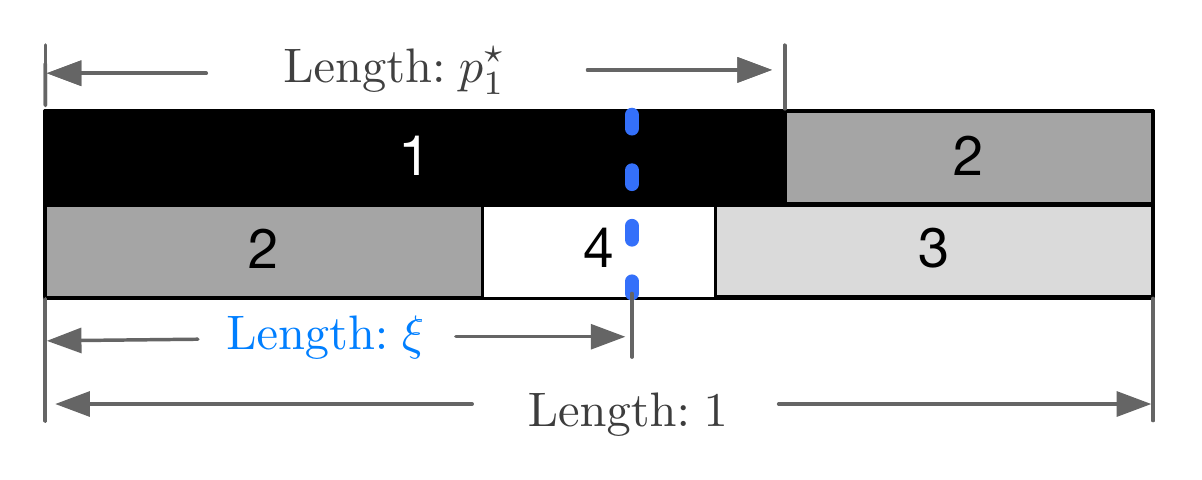}
		\label{f:HeurisSolu2}}
	\hspace{1mm}
	\subfloat[A third result.] 
	{\includegraphics[angle=0,width=0.32\textwidth]{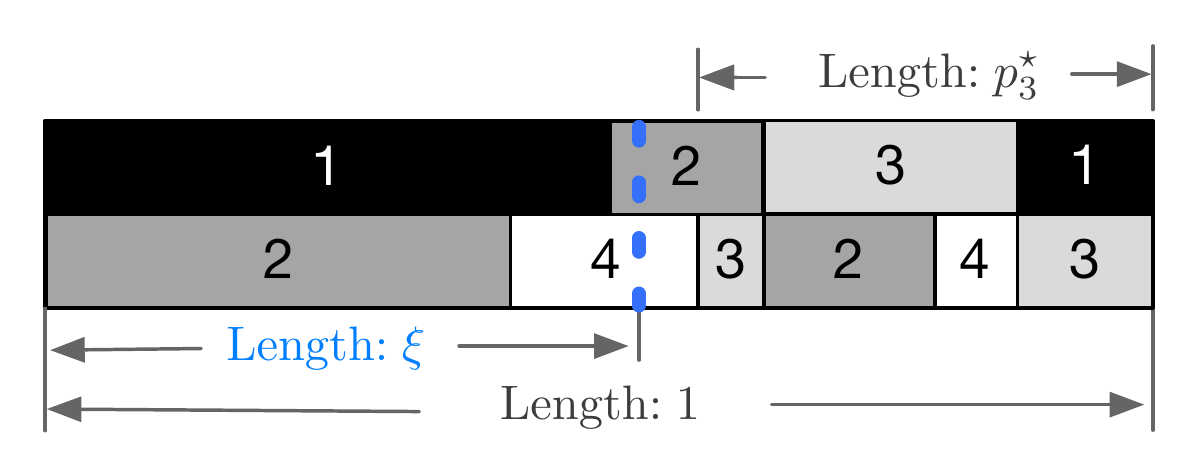}
		\label{f:HeurisSolu3}}
	\caption{Generating cache state probabilistically given $\mathbf{p}^\star$ in static probabilistic caching.} \label{f:HeurisSolu}
\end{figure}

A heuristic method for implementing the given $\mathbf{p}^\star$ is introduced in \cite{J_BBlaszczyszyn2015}. Specifically, one can generate $c$ (i.e., the cache size) rows, all with length 1, and then a block for each content, where the length of block $k$ is equal to $p^\star_k$. Next, the blocks are placed into the rows one by one. If the vacant part of a row is not long enough for a block, the rest of this block continues from the beginning of the next row. Fig.~\ref{f:HeurisSolu1} illustrates the above procedure using a total of 5 contents and a cache of size 2 as an example. In this example, the given caching probabilities for contents 1 to 4 are nonzero while the given caching probability for content 5 is zero. We represent the blocks for content~1 to content~4 using decreasing shades. 
After placing all blocks, the next step is to generate a random number $\xi$ in $[0, 1]$ based on uniform distribution and draw a vertical line (the dashed line in Fig.~\ref{f:HeurisSolu1}) at the corresponding location. Then, the blocks that the vertical line crosses indicate the contents to be cached to implement $\mathbf{p}^\star$. For example, in the case of Fig.~\ref{f:HeurisSolu1}, contents~1~and~3 will be cached.

Although the aforementioned method does not explicitly solve $\boldsymbol{\eta}$ from the given $\mathbf{p}^\star$, the probability of a chosen cache state corresponds to one solution of $\boldsymbol{\eta}^\star$. Specifically, if the dashed vertical line in Fig.~\ref{f:HeurisSolu1} sweeps across the rows from one side to the other under a constant speed, the portion of time that the vertical line crosses state $l$ is equal to $\eta^l$. However, the solution in eq.~\eqref{e:etaSolu} is non-unique in general, and probabilistic caching based on the aforementioned method corresponds to only one solution of $\boldsymbol{\eta}^\star$. If we rearrange the blocks or divide the blocks into pieces, the result may correspond to different solutions of $\boldsymbol{\eta}^\star$. Figs.~\ref{f:HeurisSolu2}~and~\ref{f:HeurisSolu3} give two examples.

As subtle as the aforementioned heuristic method is, it can be more complicated than finding a solution using eq.~\eqref{e:etaSolu} when $N_\mathrm{f}$ is large. More importantly, it is worth noting that, while the results in Figs.~\ref{f:HeurisSolu1},~\ref{f:HeurisSolu2},~or~\ref{f:HeurisSolu3} are equivalent in the case of static caching, different solutions of $\boldsymbol{\eta}$ may not be equivalent in the case of dynamic caching with content replacement. Consider the same example with 5 contents and a cache size of 2. Suppose that the optimal caching probabilities for the 5 contents found from optimization are given by $\mathbf{p}^\star = [0.9, 0.7, 0.3, 0.1, 0]^\mathrm{T}$. The matrix $\mathbf{S}$ is given as follows:
\begin{align}
\mathbf{S} = 
\begin{bmatrix}
1   &  1  &  1  &  1  &  0  &  0  &  0  &  0  &  0  &  0  \\
1   &  0  &  0  &  0  &  1  &  1  &  1  &  0  &  0  &  0 \\
0   &  1  &  0  &  0  &  1  &  0  &  0  &  1  &  1  &  0 \\
0   &  0  &  1  &  0  &  0  &  1  &  0  &  1  &  0  &  1 \\
0   &  0  &  0  &  1  &  0  &  0  &  1  &  0  &  1  &  1
\end{bmatrix} \nonumber
\vspace{-5mm}
\end{align}
Then, it can be shown that 
\begin{displaymath}
\boldsymbol{\eta}_1 = [0.6, 0.2, 0.1, 0, 0.1, 0, 0, 0, 0, 0]^\mathrm{T},
\end{displaymath} 
which corresponds to the case similar to Fig.~\ref{f:HeurisSolu2} and
\begin{displaymath}
\boldsymbol{\eta}_2 = [0.62, 0.22, 0.06, 0, 0.06, 0.02, 0, 0.02, 0, 0]^\mathrm{T},
\end{displaymath} 
which corresponds to the case similar to Fig.~\ref{f:HeurisSolu3}, can both implement $\mathbf{p}^\star = [0.9, 0.7, 0.3, 0.1, 0]^\mathrm{T}$. Therefore, for static caching, content placement generated from $\boldsymbol{\eta}_1$ and $\boldsymbol{\eta}_2$ are equivalent. However, for dynamic caching, dynamic content replacement based on $\boldsymbol{\eta}_1$ can lead to a less number of cache replacements than that based on $\boldsymbol{\eta}_2$. This is because $\boldsymbol{\eta}_1$ has less nonzero elements, corresponding to fewer cache states. For example, dynamic implementation of $\mathbf{p}^\star$ based on $\boldsymbol{\eta}_2$ may cause a replacement of content 3 by content 4 when the cached contents are $\{2, 3\}$ and a replacement of content 1 by content 3 when the cached contents are $\{1, 4\}$. However, such replacements will not happen if dynamic caching based on $\boldsymbol{\eta}_1$ is used because $\boldsymbol{\eta}_1$ assigns zero probability to the states that cache $\{2, 4\}$ and $\{3,4\}$.


Note that the choice of $\boldsymbol{\eta}^\star$ based on the optimal content caching probability $\mathbf{p}^\star$ can determine the initial content placement. Accordingly, the initial contents stored in the target cache at the beginning of the considered time window are pre-fetched based on $\boldsymbol{\eta}^\star$, which occurs at the beginning of the time window. Then, for dynamic caching, the target cache updates its cached contents as it receives content requests, as studied in the next section.



\section{Probabilistic Content Replacement: the Markov Chain of Content Replacement}


In classic dynamic caching such as LRU, replacements are usually determined solely by the content request statistics. Such replacements provide adaptivity to time-varying content popularity without any \textit{a priori} information. However, in our considered scenario, in which average content popularity $\{\varphi_k\}_{k \in \mathcal{F}}$ and target content caching probabilities $\{p_k^\star\}_{\forall k}$ are available, using classic dynamic caching while neglecting the average content popularity can lead to a waste of using information. Therefore, we aim to design dynamic probabilistic caching that can exploit the average content popularity $\{\varphi_k\}_{k \in \mathcal{F}}$ to adapt to the content popularity. A logical requirement is that the resulting caching probability should converge to the optimal content placement policy $\boldsymbol{\eta}^\star$ based on the average content popularity in the case of time-invariant content popularity. This section addresses two questions regarding the design of such dynamic caching: a) what should be the probability of accepting a new content into the cache? and b) which existing content should be replaced, and with what probability, if the new content is accepted. In the rest of this section, we assume that the target cache makes replacement decisions locally, i.e., without needing to check with other caches.

\subsection{The content replacement Markov chain}


The content replacement process at the target cache can be modeled using a Markov chain. Note that the idea of using Markov chain and related tools in designing content replacement policy can be found in the literature. For example, using a Markov decision process, Bahat and Makowski proved that the optimal content replacement policy is a Markov stationary policy under the independent reference model~\cite{C_OBahat2003}. In their recent work, Shukla and Abouzeid modeled content replacement using a Markov decision process and found the optimal content retention time to jointly minimize content retrieval delay and cache wearout~\cite{J_SShukla2017}. Our focus here, however, is the design of content replacement that can converge to a target stationary point in the case when the instantaneous content popularity is time-invariant and adapt to the variations in the case when the instantaneous content popularity is time-varying.

In the content replacement Markov chain, a state transition may happen when a requested content is not in the cache. For state $m$, denote the set of states that state $m$ can transit into by replacing one cached content with content $k$ and the entire set of states that state $m$ can transit into by replacing one cached content with any other content as $\mathcal{H}_{m}^k$ and $\mathcal{H}_{m}$, respectively. It is not difficult to see that state $m$ and any state in $\mathcal{H}_{m}$ must differ in one and only one cached content. Moreover, the following results hold:
\begin{align}\label{e:NeighborConds}
\mathcal{H}_{m} =\mathop{\bigcup}\limits_{k \in \mathcal{F}\backslash \mathcal{G}^m} \mathcal{H}_{m}^k, \;\; |\mathcal{H}_{m}^k| = c, \;\; |\mathcal{H}_{m}| = c (N_\mathrm{f} - c). 
\end{align}


Consider a pair of neighbor states $m$ and $m^\prime\in \mathcal{H}_{m}^k$. The question that arises is: with what probability should the cache transition into state $m^\prime$ given that content $k$ is requested while the cache is currently in state $m$? Denote this conditional probability as $\tau_{m^\prime, m}$. Then, the design of dynamic probabilistic caching boils down to finding $\tau_{m^\prime, m}$ for every $m$ and $m^\prime\in \mathcal{H}_{m}^k$ where $k \in \mathcal{F}\backslash \mathcal{G}^m$.

The overall state transition probability matrix of the content replacement Markov chain is determined by two sets of probabilities: the content request probabilities and $\{\tau_{m^\prime, m}\}$. Since the instantaneous content request probabilities are unknown, we can only exploit the average content request probabilities $\{\varphi_k\}_{k \in \mathcal{F}}$. Therefore, the design of dynamic content replacement uses $\{\varphi_k\}_{k \in \mathcal{F}}$ instead of the instantaneous content request probabilities. A discussion on the effect of time-varying content popularity will be given later.

Denote the overall state transition probability matrix by $\boldsymbol{\Theta}$. The element at the $m$th column and the $m^\prime$th row of $\boldsymbol{\Theta}$ represents the probability that state $m$ transitions into state $m^\prime$. Then, $\boldsymbol{\Theta}$ can be written as the summation of content-specific conditional state transition matrices as follows:
\begin{align}
\boldsymbol{\Theta} =\sum\limits_{k \in \mathcal{F}} \varphi_k \boldsymbol{\Theta}_k. 
\end{align} 	
where $\boldsymbol{\Theta}_k$ is the conditional state transition probability matrix given that content $k$ is being requested. It can be seen that $\boldsymbol{\Theta}$ and $\boldsymbol{\Theta}_k, \forall k$ have many zero elements because $\boldsymbol{\Theta}(m, m^\prime)$ and $\boldsymbol{\Theta}(m^\prime, m)$ are both zero if $m^\prime\notin \mathcal{H}_{m}$, and $\boldsymbol{\Theta}_k(m, m^\prime)$ and $\boldsymbol{\Theta}_k(m^\prime, m)$ are both zero if $m^\prime\notin \mathcal{H}_{m}^k$. Specifically, based on the conditions in eq.~\eqref{e:NeighborConds}, each column of $\boldsymbol{\Theta}$ only has $c (N_\mathrm{f} - c) + 1$ nonzero elements with one on and the rest off the main diagonal. For a column in $\boldsymbol{\Theta}_k$, two cases are possible. If state $m$ caches content $k$, then the diagonal element $\boldsymbol{\Theta}_k(m, m)$ is the only nonzero element in the $m$th column. Otherwise, the $m$th column has $c$ nonzero elements off the main diagonal.

\begin{figure}[tt]
	\centering {\includegraphics[width=0.38\textwidth]{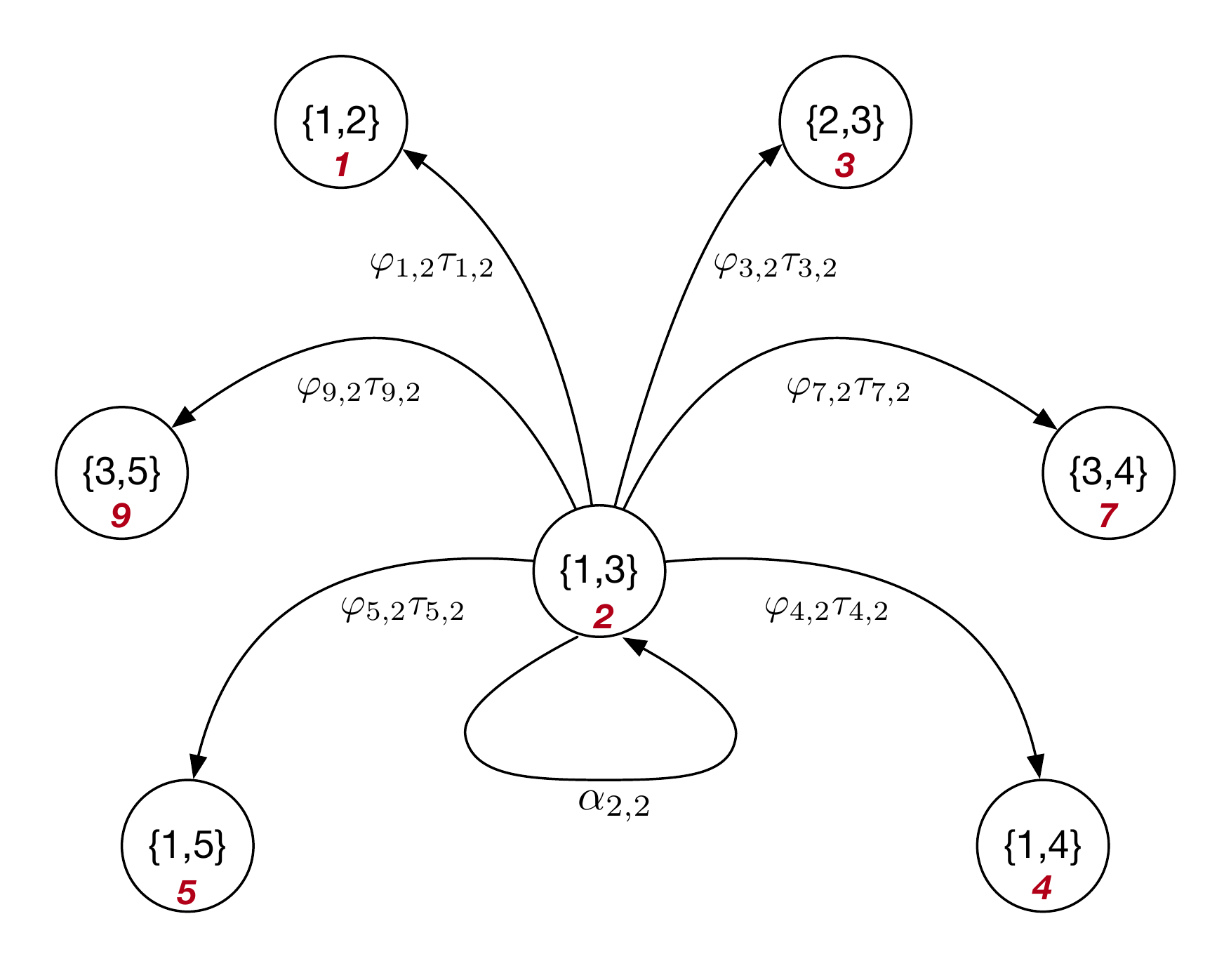}}
	\vspace{-3mm}
	\caption{An illustration of the variables in the state transition using the example of state 2, where $N_\mathrm{f} = 5$ and $c= 2$. Each circle represents a state, the bracketed numbers in the center represent the cached contents in that state, and the italic number at the bottom represents the state ID. }\label{f:illuStateTrans} 
\end{figure}

In order to highlight the state transition, the content request probability $\varphi_k$ at the target cache is alternatively denoted by $\varphi_{m^\prime, m}$ if $m^\prime\in \mathcal{H}_{m}^k$, i.e., if content $k$ must be requested for the cache to transition from state $m$ into state $m^\prime$. Denote the $m$th element on the main diagonal of $\boldsymbol{\Theta}$ by $\alpha_{ m, m}$. Fig.~\ref{f:illuStateTrans} illustrates the state transitions. In the illustrated example, $N_\mathrm{f} = 5$ and $c= 2$, and therefore $c (N_\mathrm{f} - c) + 1 = 7$. As a result, each state can transition into itself and 6 other states.

Given the above notations, the overall transition probability matrix $\boldsymbol{\Theta}$ corresponding to the content replacement Markov chain for the target cache can be written as:
\begin{align}\label{e:TransMatrixTi} 
\boldsymbol{\Theta} \!=\!
\scalebox{0.95}{$\kbordermatrix{\!\!
	~   &  1  & \!\!\!\cdots\!\!  &  m  &  \!\!\!\cdots \!\! &  n  \cr
	1\!\!   & \!\! \alpha_{ 1, 1} &  \!\! \scalebox{.75}{$\cdots$} \!\! &   \varphi_{ 1, m}\tau_{1, m} &  \!\! \scalebox{.75}{$\cdots$} \!\!  & \varphi_{ 1, n}\tau_{1, n}\!\!  \cr
	\scalebox{.75}{$\vdots$}\!\! & \scalebox{.75}{$\vdots$}    &  \!\! \scalebox{.75}{$\ddots$}\!\!   &  \scalebox{.75}{$\vdots$}  &   & \scalebox{.75}{$\vdots$} \cr
	m\!\! &  \!\!\varphi_{ m, 1}\tau_{m, 1}  &   &  \alpha_{m, m}   &    & \varphi_{ m, n}\tau_{m, n} \!\! \cr
	\scalebox{.75}{$\vdots$}\!\! & \scalebox{.75}{$\vdots$}     &   & \scalebox{.75}{$\vdots$}  & \!\!\scalebox{.75}{$\ddots$}\!\!   & \scalebox{.75}{$\vdots$} \cr
	n\!\!  & \!\!  \varphi_{ n, 1}\tau_{n, 1}   &  \!\!\scalebox{.75}{$\cdots$}\!\!   & \varphi_{ n, m}\tau_{ n, m}  & \!\!\scalebox{.75}{$\cdots$}\!\!  & \alpha_{ n, n}  \!\!\cr}$},
\end{align}
where $\varphi_{ m^\prime, m} \in [0, 1]$ and $\tau_{ m^\prime, m} \in [0, 1], \forall m^\prime \in \mathcal{H}_{m}, \forall m$. The diagonal element $\alpha_{m, m}$ in \eqref{e:TransMatrixTi} can be found as follows: 
\begin{align}\label{e:alpha}
\alpha_{ m, m} \!=\!   \sum\limits_{k \in \mathcal{G}^m} \varphi_{k}  \! + \!  \sum\limits_{m^\prime \in \mathcal{H}_{ m}} \!\!  \varphi_{m^\prime, m} (1  -  \tau_{ m^\prime, m} ). 
\end{align}
The first item in eq.~\eqref{e:alpha} represents the probability that a content currently in the cache is requested, while the second item represents the probability that a content not in the cache is requested (and downloaded) but not accepted into the cached (i.e., no replacement occurred).


In Section~\ref{s:ProbContentPlace}, we have investigated the problem of implementing target content caching probabilities $\mathbf{p}^\star$ by finding the corresponding state caching probabilities $\boldsymbol{\eta}^\star$. Given $\boldsymbol{\eta}^\star$, the design of dynamic probabilistic caching so that the content caching probabilities converge to $\mathbf{p}^\star$ in the case of time-invariant content popularity is equivalent to finding the transition matrix $\boldsymbol{\Theta}$ such that
\begin{align}\label{e:TransSteady}
\boldsymbol{\Theta} \boldsymbol{\eta}^\star = \boldsymbol{\eta}^\star. 
\end{align} 
If the elements of $\boldsymbol{\Theta}$ could be arbitrarily chosen in the range of $[0,1]$, the problem can be solved with existing methods, e.g., the Metropolis-Hastings Algorithm \cite{CPRobert2004}. However, as can be seen from  eq.~\eqref{e:TransMatrixTi}, there are additional constraints on the elements of $\boldsymbol{\Theta}$. First, each off-diagonal element is a product of two items, and the $(m^\prime, m)$th element should be bounded by $\varphi_{m^\prime, m}$. Second, the summation of multiple elements in the same column should also be bounded, i.e., 
\begin{align}
 \sum\limits_{\forall m^\prime \in \mathcal{H}_{m}^k} \varphi_{m^\prime, m}\tau_{m^\prime, m} \leq \varphi_k, \forall k \in \mathcal{G}^m, \forall m.   
\end{align}
Consequently, the Metropolis-Hastings Algorithm cannot be applied to the considered problem. In the next section, an approach is proposed to construct an irreducible and ergodic Markov chain by designing the matrix $\boldsymbol{\Theta}$ for the target cache so that $\boldsymbol{\eta}^\star$ is the unique steady state.  


\subsection{Generating the state transition matrix $\boldsymbol{\Theta}$}

Without loss of generality, it is assumed that the elements of $\boldsymbol{\eta}^\star$ are non-zero and arranged in a non-increasing order, i.e., $\eta^q \geq \eta^l$ if $q <l$. An extension to the case when  $\eta^{j}$ is zero for some $j$ is straightforward. 

When a content not in the cache is requested, a replacement may or may not happen. In order to control this factor in our design, we introduce parameters $\{\omega_{k,  m^\prime, m}\}$ to represent the upper-limit on state transition probabilities. Specifically, for any given $m$ and $m^\prime \in \mathcal{H}_{m}^k$, the parameter $\omega_{k,  m^\prime, m}$ represents the upper limit that state $m$ transits into state $m^\prime$ given that content $k$ is requested. As a result, the following condition must be satisfied 
\begin{align}
\sum\limits_{m^\prime \in \mathcal{H}_{m}^k} \omega_{k, m^\prime, m} \leq 1, \forall m, \forall k.
\end{align}  
If strict equality holds in the above condition, then content $k$ is always accepted into the cache when the cache state is $m$. Otherwise, content $k$ may not be accepted into the cache even after it is requested and downloaded.  

\begin{figure}[!t]
	\centering 
	\subfloat[Step~1: original input sequence.]
	{\includegraphics[angle=0,width=0.45\textwidth, scale=0.9]{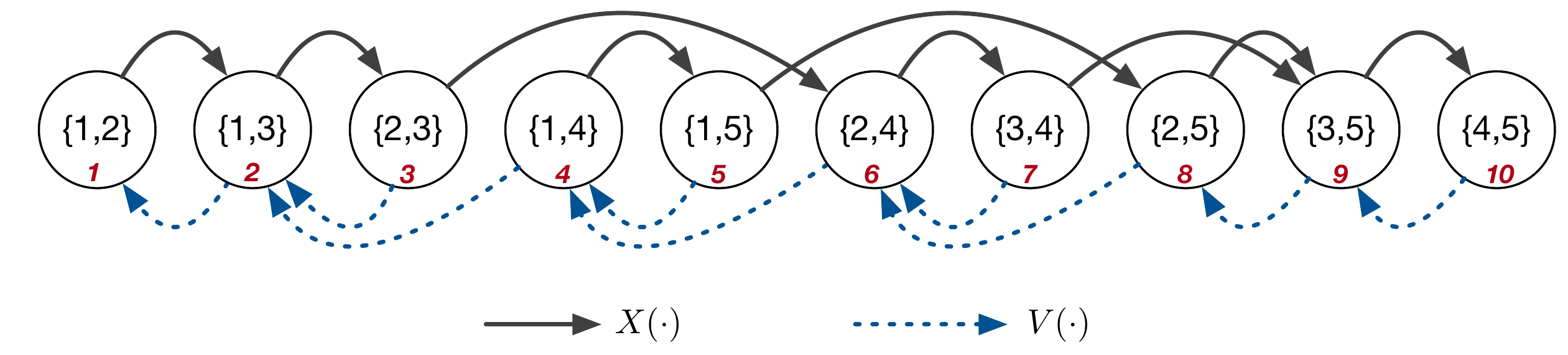}
		\label{f:VX0}}
	\hspace{1mm} 
	\subfloat[Step 2: removing a state from the original sequence.] 
	{\includegraphics[angle=0,width=0.41\textwidth, scale=0.9]{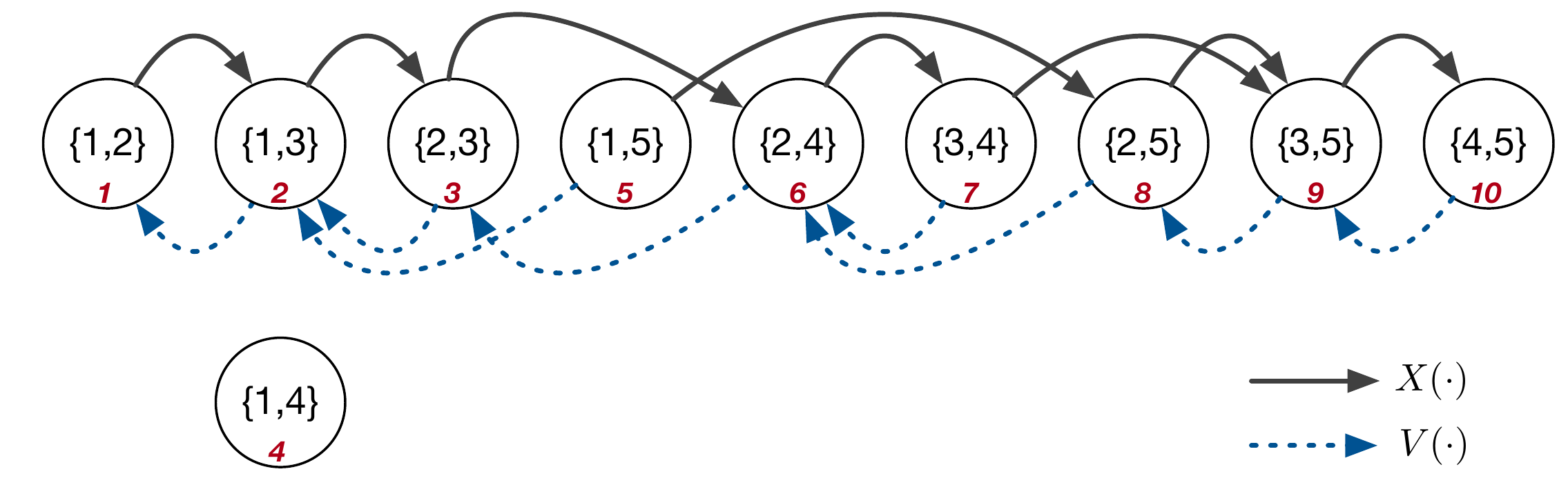}
		\label{f:VX1}}
	\hspace{1mm} 
	\subfloat[Step 3: removing 2nd state from the original sequence.]
	{\includegraphics[angle=0,width=0.37\textwidth, scale=0.7]{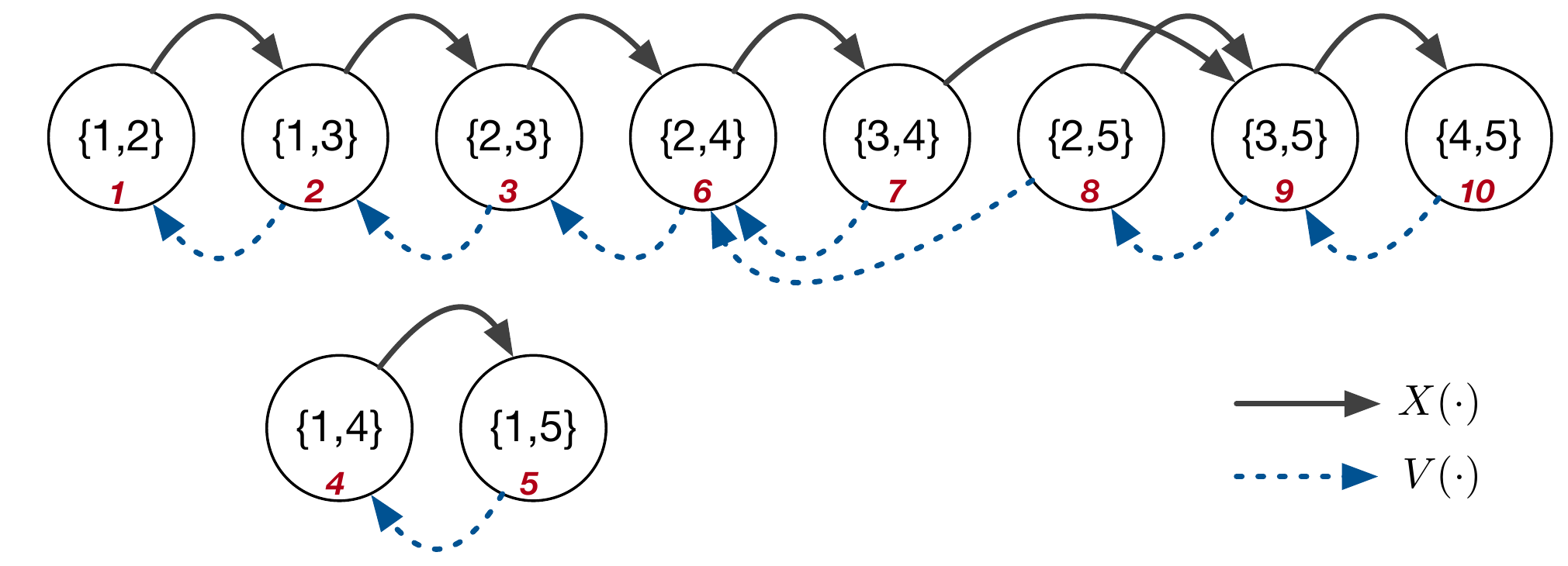}
		\label{f:VX2}}
	\hspace{1mm} 
	\subfloat[Step 4: the final ordered sequences.]
	{\includegraphics[angle=0,width=0.32\textwidth, scale=0.5]{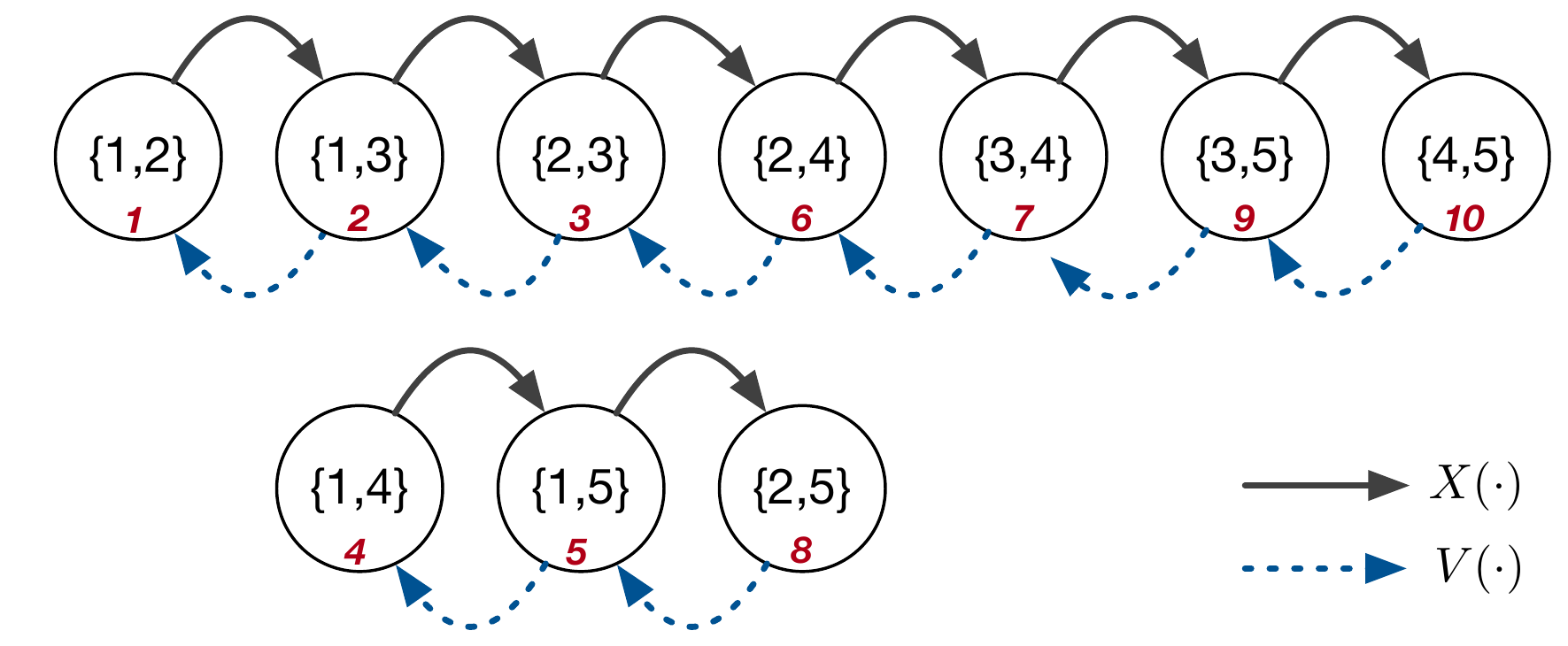}
		\label{f:VX3}}
	\caption{An illustration of using $V(m)$ and $X(m)$ to group states into ordered sequences. The states satisfy $\eta_1^\star \geq \eta_2^\star \geq ... \geq \eta_{10}^\star$.} \label{f:IlluVx} 
\end{figure}

The next step is to determine which state transitions could happen and with what probabilities. Recall that the elements of $\boldsymbol{\eta}^\star$ are arranged in non-increasing order. Note that adjacent states, e.g., state $m$ and state $m+1$, may not be neighbor states using such an order. For any given  $m$, define the functions $V(m)$ for $m \in \{2, \dots, n\}$ and $X(m)$ for $m \in \{1, \dots, n -1\}$, both mapping from a state to one of its neighbor states as follows
\begin{subequations}
\begin{align}
V(m) = \mathop{\text{arg}\min}\limits_{\hat{m} \in \mathcal{H}_{m} }\{\eta^{\hat{m}} | \eta^{\hat{m}} \geq \eta^{m} \}, \\
X(m) = \mathop{\text{arg}\max}\limits_{\check{m} \in \mathcal{H}_{m}}\{\eta^{\check{m}} | \eta^{\check{m}} \leq \eta^{m}\}.
\end{align}
\end{subequations} 
An illustration of $V(\cdot)$ and $X(\cdot)$ when $N_\mathrm{f} = 5$ and $c = 2$ is given in Fig.~\ref{f:VX0}. Using $V(m)$ and $X(m)$ on state $m$, two cases are possible. 
\begin{itemize}
\item $X(m) = m+1$ and $V(m+1) = m$: In such case, states $m$ and $m+1$ are both adjacent and neighbor states (e.g., states 1~and~2 in Fig.~\ref{f:VX0}). 
\item $X(m) \neq m + 1$ and $V(m+1) \neq m$: In such case, $m$ and $m+1$ are adjacent but not neighbor states (e.g., states 3~and~4 in Fig.~\ref{f:VX0}).
\end{itemize}

To facilitate the design of state transitions, we organize the states into several sequences so that: 1).  $\eta_m^\star$ in each sequence is sorted in a non-increasing order; and 2). adjacent states in the same sequence are always neighbors. Denote the original sequence of all states by $\mathsf{S}_0$. Using the procedure in Algorithm~\ref{a:Sequence} repeatedly (by setting the output $\mathsf{L}$ as the input sequence $\mathsf{S}_0$ of the next run until the output $\mathsf{L}$ is empty), the above-mentioned ordered sequences can be obtained. The procedure is illustrated in 4 steps in Fig.\ref{f:IlluVx}.

 \begin{algorithm}[tt!]
	\caption{Generating ordered sequence of neighbor states}\label{a:Sequence}
	\begin{algorithmic}[1]
		\renewcommand{\algorithmicrequire}{\textbf{Input:}} 
		\renewcommand{\algorithmicensure}{\textbf{Output:}}
		\REQUIRE $\mathsf{S}_0$
		\ENSURE $\mathsf{S}, \mathsf{L}$ 	\\ 
		\hspace{-6mm}\textit{Initialization}: $\mathsf{S} = \mathsf{S}_0$; L set to an empty sequence. 
		\FOR {State $m = 2$ to $n$}
		\IF{$X(V(m)) \neq m$}
		\STATE Mark $V(m)$ as a potential connection point; \\
			   Remove $m$ from sequence $\mathsf{S}$; \\
			   Add $m$ to the end of sequence $\mathsf{L}$; \\
		\ELSIF{$V(X(m)) \neq m$}
		\STATE Mark $X(m)$ as a potential connection point; 
		\ENDIF
		\ENDFOR
		\RETURN $\mathsf{S}, \mathsf{L}$	
	\end{algorithmic}
\end{algorithm}

\begin{figure}[!t]
	\centering \subfloat[Ordered state sequence $\mathsf{S}_1$ consists of the 7 states in the first row, and $\mathsf{S}_2$ consists of the 3 states in the second row. The states~2~and~9 are the branch and merge points of $\mathsf{S}_2$, respectively. The dashed transition links are created by Steps~2~and~3 of Algorithm~\ref{a:Tgen}.]
	{\includegraphics[angle=0,width=0.48\textwidth]{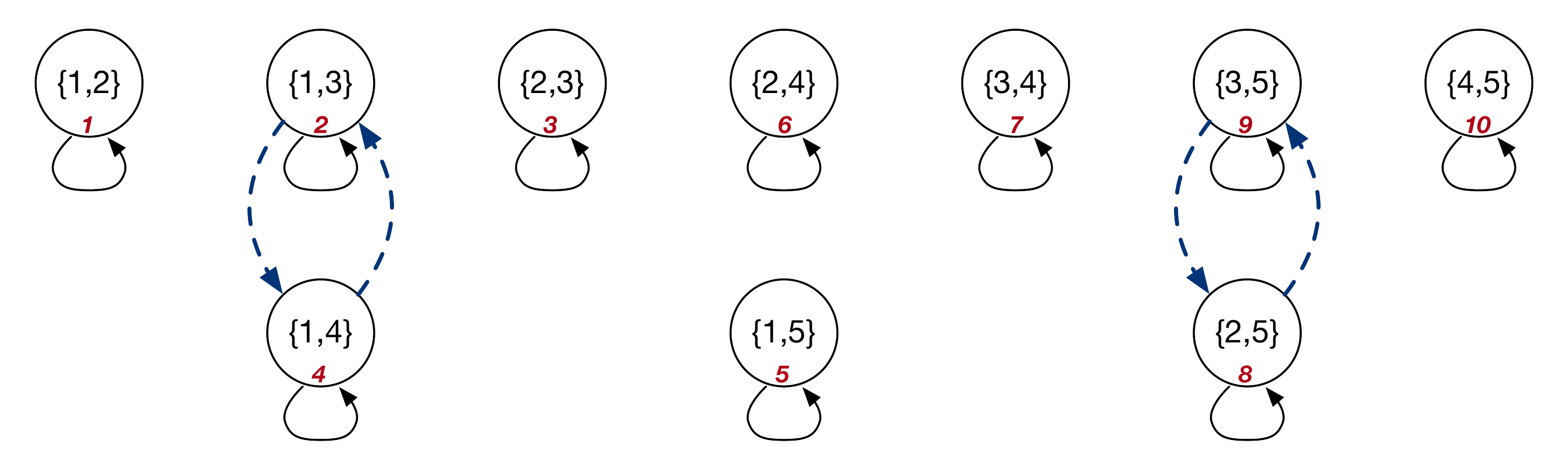}
		\label{f:AlgoIlluStep1}}
	\hspace{1mm} 
	\subfloat[The dashed transition links are created by the update on $\boldsymbol{\Theta}$ by Steps~4~to~6 of Algorithm~\ref{a:Tgen}.] 
	{\includegraphics[angle=0,width=0.48\textwidth]{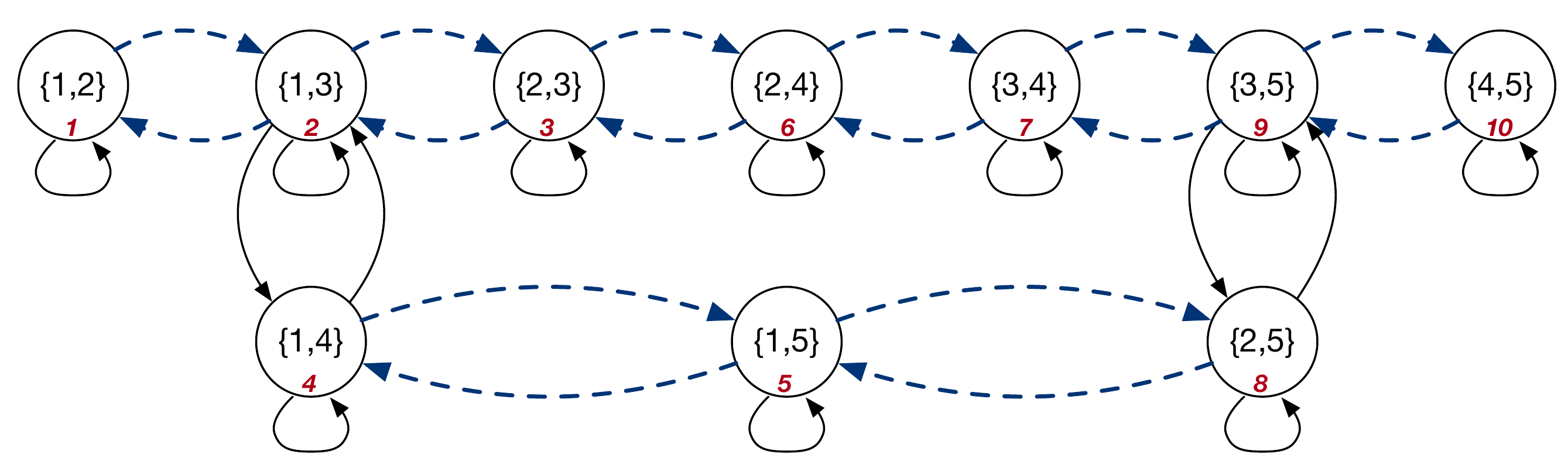}
		\label{f:AlgoIlluStep2}}
	\hspace{1mm} 
	\subfloat[The Markov chain created after $\boldsymbol{\Theta}$ is refined by Algorithm~\ref{a:RefineT}. To reduce the number of connections, bi-direcitional links are used.]
	{\includegraphics[angle=0,width=0.48\textwidth]{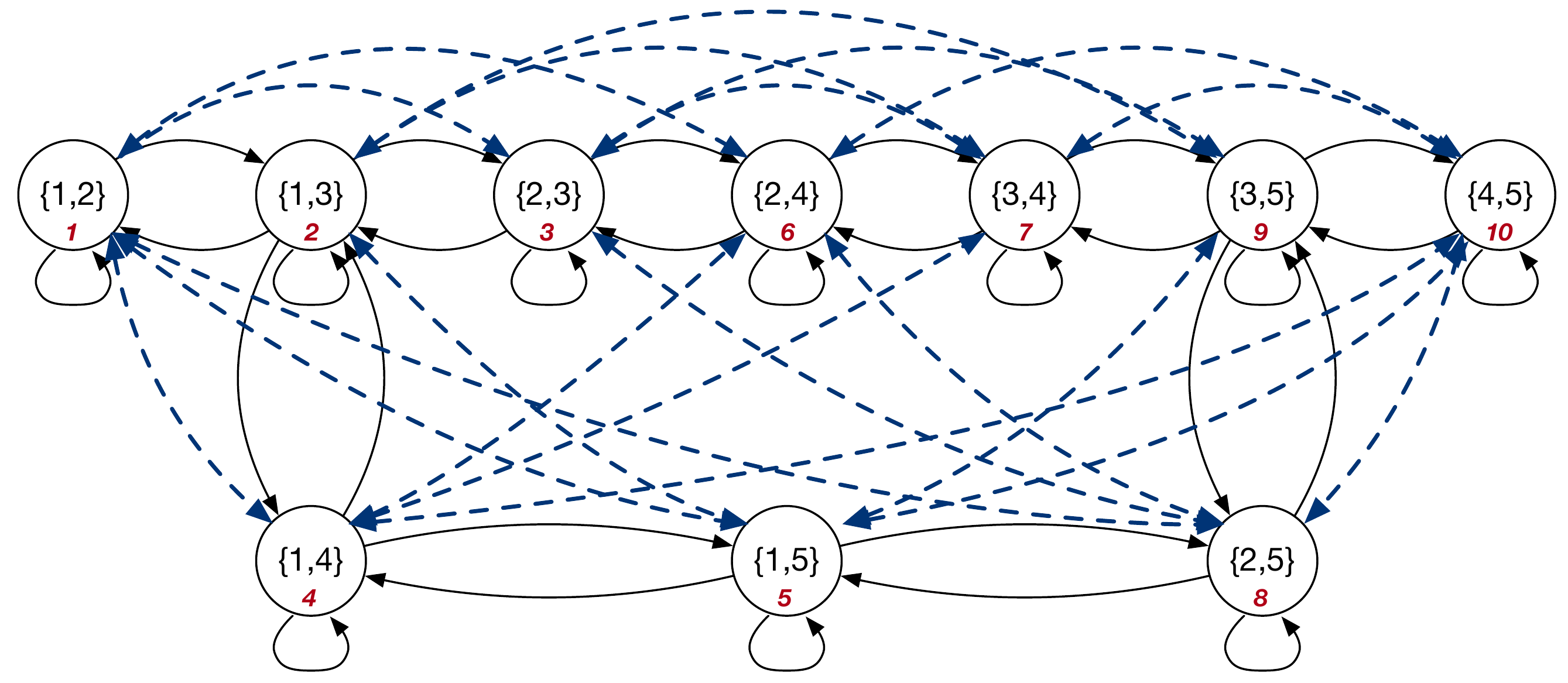}
		\label{f:AlgoIlluStep3}}
	\caption{An illustration on generating and refining the underlying Markov chain by updating $\boldsymbol{\Theta}$ using Algorithms~\ref{a:Tgen}~and~\ref{a:RefineT}.} 
\end{figure}

The connection points identified in Step~3 and Step~5 of Algorithm~\ref{a:Sequence} are the points where the next sequence of states may connect with the current sequence in the Markov chain. We refer to the points where the first state and the last state of the next sequence can connect to as branch and merge points, respectively. Determine one branch point and one merge point for each sequence is not difficult and neglected here. Denote the $l$th sequence of states as $\mathsf{S}_l$ and the length of $\mathsf{S}_l$ as $K_l$. Denote the $k$th state, the branch point, and the merge point of sequence $\mathsf{S}_l$ as $\mathsf{S}_l(k)$, $B(l)$, and $M(l)$, respectively. This is illustrated in Fig.~\ref{f:AlgoIlluStep1}.

Given the ordered sequences, we next determine the state transition probabilities iteratively, considering one pair of states in each iteration. Initialize the state transition probability matrix $\boldsymbol{\Theta}$ to be $\mathbf{I}$. The basic procedure for updating the state transition probabilities for a pair of neighbor states $m$ and $m$ such that $m^\prime = \mathcal{H}_{i, m}^k$ and $m = \mathcal{H}_{m^\prime}^{k^\prime}$ is given as follows:
\begin{subequations}\label{e:BasicProcedure}
	\begin{align}
	&\delta  = \min \bigg\{\omega_{k, m^\prime, m}  \varphi_{ m^\prime, m}, \; \omega_{k^\prime, m, m^\prime}  \varphi_{ m, m^\prime} \frac{\eta^{m^\prime}}{\eta^m}\bigg\},   \label{e:BasicProcedure1} \\
	&\boldsymbol{\Theta}(m^\prime, m) = \delta, \label{e:BasicProcedure2} \\
	&\boldsymbol{\Theta}(m, m^\prime) = \delta  \frac{\eta^m}{\eta^{m^\prime}},  \label{e:BasicProcedure3}  \\
	&\boldsymbol{\Theta}(m^\prime, m^\prime) \Leftarrow \boldsymbol{\Theta}(m^\prime, m^\prime) - \delta\frac{\eta^m}{\eta^{m^\prime}},  \label{e:BasicProcedure4}  \\  
	&\boldsymbol{\Theta}(m, m) \Leftarrow \boldsymbol{\Theta}(m, m) - \delta,    \label{e:BasicProcedure6}
	\end{align} 
\end{subequations}
where $\Leftarrow$ represents the operation of assigning the value of the right-hand side expression to the left-hand side. The above procedure guarantees that, assuming $\boldsymbol{\Theta} \boldsymbol{\eta}^\star = \boldsymbol{\eta}^\star$ before the update, the equality still holds after updating the transition probabilities for states $m$ and $m^\prime$.

 \begin{algorithm}[tt!]
	\caption{Generating the Stochastic Matrix $\boldsymbol{\Theta}$}\label{a:Tgen}
	\begin{algorithmic}[1]
		\renewcommand{\algorithmicrequire}{\textbf{Input:}} 
		\renewcommand{\algorithmicensure}{\textbf{Output:}}
		\REQUIRE  $\boldsymbol{\eta}^\star$, $\{\varphi_{m, m^\prime}\}_{\forall m, \forall m^\prime}$, $\{\mathsf{S}_l\}$,  $\{B(l)\}$, $\{M(l)\}$ 
		\ENSURE  $\boldsymbol{\Theta}$ 	\\ 
		\hspace{-6mm}\textit{Initialization}: Set $\boldsymbol{\Theta} = \mathbf{I}$. 
		\FOR {each sequence $l$}
		\STATE Run the procedure \eqref{e:BasicProcedure1}-\eqref{e:BasicProcedure6} with $m = B(l), m^\prime = \mathsf{S}_l(1)$.
		\STATE Run the procedure \eqref{e:BasicProcedure1}-\eqref{e:BasicProcedure6} with $m = M(l), m^\prime = \mathsf{S}_l(K_l)$. 
		\FOR {State $q = K_l$ to 2}
		\STATE Run the procedure \eqref{e:BasicProcedure1}-\eqref{e:BasicProcedure6} with $m^\prime = \mathsf{S}_l(q)$ and $m = \mathsf{S}_l(q-1)$. 
		\ENDFOR
		\ENDFOR
		\RETURN $\boldsymbol{\Theta}$
	\end{algorithmic}
\end{algorithm}

Based on the above definitions and procedures, Algorithm~\ref{a:Tgen} is proposed to generate a basic transition probability matrix $\boldsymbol{\Theta}$ that has the steady state $\boldsymbol{\eta}^\star$. An illustration of Algorithm~\ref{a:Tgen} is given in Figs.~\ref{f:AlgoIlluStep1}~and~\ref{f:AlgoIlluStep2}. Steps~2~and~3 of Algorithm~\ref{a:Tgen} ``connect'' the ordered sequences found using Algorithm~\ref{a:Sequence} by generating the state transition probabilities for the connection points, e.g., states~2~and~4 and states~8~and~9 as illustrated in Fig.~\ref{f:AlgoIlluStep1}. Then, Steps~4~to~6 of Algorithm~\ref{a:Tgen} generate state transitions only between adjacent and neighbor states within each ordered sequence of states. This is illustrated in Fig.~\ref{f:AlgoIlluStep2}, which shows a basic irreducible Markov chain at the output of Algorithm~\ref{a:Tgen}. The generated Markov chain satisfies $\boldsymbol{\Theta} \boldsymbol{\eta}^\star = \boldsymbol{\eta}^\star$ but most of the off-diagonal elements in $\boldsymbol{\Theta}$ are 0. As a result, the mixing time can be long. In order to reduce the mixing time, we use a refinement procedure to connect more states, which is given in Algorithm~\ref{a:RefineT} and illustrated in Fig.~\ref{f:AlgoIlluStep3}. The refinement in Algorithm~\ref{a:RefineT} is designed based on the fact that the mixing time of the Markov chain is determined by the second largest eigenvalue of the transition matrix \cite{SBoydSIAM2004}. Simulation examples in Section~\ref{s:Simu} will demonstrate the performance of the refinement. Detailed analysis, however, is beyond the scope of this work.


 \begin{algorithm}[hb!]
	\caption{Refining the matrix $\boldsymbol{\Theta}$}\label{a:RefineT}
	\begin{algorithmic}[1]
		\renewcommand{\algorithmicrequire}{\textbf{Input:}} 
		\renewcommand{\algorithmicensure}{\textbf{Output:}}
		\REQUIRE $\boldsymbol{\Theta}$, $\{\varphi_{m, m^\prime}\}_{\forall m, \forall m^\prime}$
		\ENSURE  $\boldsymbol{\Theta}$ 	\\ 
		\FOR {State $m = 1$ to $n$}
		\FOR {State $m^\prime = m+ 1$ to $n$}
		\IF { $m^\prime \in \mathcal{H}_{m}$ and $\boldsymbol{\Theta}(m^\prime, m)=0$ }
		\STATE Run the procedure in \eqref{e:BasicProcedure1}-\eqref{e:BasicProcedure6}
		\ENDIF
		\ENDFOR
		\ENDFOR
		\RETURN $\boldsymbol{\Theta}$	
	\end{algorithmic}
\end{algorithm}


\begin{theorem}\label{t:PropTi}
	In the case when the instantaneous content popularity converges to the average content popularity, the matrix $\boldsymbol{\Theta}$ generated by Algorithm~\ref{a:Tgen} has the following properties:
	\begin{itemize}
		\item $\boldsymbol{\Theta}$ is a valid state transition matrix;
		\item $\boldsymbol{\Theta}$ guarantees that $\tau_{ m, m^\prime} \in [0,1]$ for all $m$ and $m^\prime \neq m$;
		\item $\boldsymbol{\Theta}$ satisfies eq.~\eqref{e:TransSteady};
		\item The underlying Markov chain specified by $\boldsymbol{\Theta}$ has a unique steady state which is $\boldsymbol{\eta}^\star$.
	\end{itemize}
\end{theorem}

\emph{Proof}: See Appendix~B.  

Based on Theorem~\ref{t:PropTi}, the underlying Markov chain can converge in distribution to $\boldsymbol{\eta}^\star$ when the instantaneous content popularity is constant. Accordingly, the content caching probabilities converge to the optimal caching probabilities $\mathbf{p}^\star$. When the instantaneous content popularity is time-varying, dynamic caching based on the designed state transitions may adapt to varying content popularity. Nevertheless, it should be noted that the adaptivity depends on the specific content request statistics, and therefore neither dynamic or static caching can claim to outperform the other in all cases. We will demonstrate and discuss this in Section~\ref{s:Simu}.


In the case of time-varying content popularity, assume that there are $N_\mathrm{r}$ requests in the considered time window. Denote the instantaneous content popularity at the $q$th request by $\{\varphi_k^{(q)}\}_{\forall k \in \mathcal{F}}$, where $q\in \{1, \dots, N_\mathrm{r}\}$. The average content popularity is specified by $\{\varphi_k\}_{\forall k \in \mathcal{F}}$. The instantaneous state transition matrix $\boldsymbol{\Theta}^{(q)}$ is time-varying as a result of the time-varying instantaneous content popularity.

\begin{lemma}\label{l:Convergence}
	Under time-varying content popularity, the overall state transition matrix averaged over the $N_\mathrm{r}$ requests, denoted by $\bar{\boldsymbol{\Theta}}$, is equal to the matrix $\boldsymbol{\Theta}$ generated by Algorithm~\ref{a:Tgen}.
\end{lemma}
\textit{Proof}:  See Appendix~C.  

It should be noted that the result $\bar{\boldsymbol{\Theta}} = \boldsymbol{\Theta}$ in Lemma~\ref{l:Convergence} does not imply that the average content caching probabilities are equal to the target content caching probabilities when the instantaneous content popularity is time-varying. To the best of our knowledge, designing dynamic caching which can achieve target average content caching probabilities without knowing the instantaneous content request probabilities is impossible. After all, the ability of adapting to content popularity implies that the resulting content caching probabilities depend on the content request statistics (not just the average content popularity).

\subsection{Discussion on the scalability}

The proposed design involves finding the transition probabilities for each cache state, and the overall number of cache states, i.e., $\binom{N_\mathrm{f}}{c}$, can be prohibitively large in practice. Nevertheless, we can limit the number of states to be considered. Next, we provide some methods for reducing the number of states when the number of contents is large. Section.~\ref{s:Simu} will use a numerical example to demonstrate the simplification of the proposed dynamic caching and the resulting performance.

\emph{On the content level}: Although the number of contents can be large, the number of ``popular'' contents which are worth caching can be small. The study in \cite{NCarlssonTPDS2017} shows that a large portion of YouTube videos ($> 70\%$) are requested only once from an edge network. It follows that a significant portion of contents will be assigned with a caching probability of zero. This can be seen in the example illustrating Fig.~\ref{f:HeurisSolu2} in Section~\ref{ss:ProbCaching}. The simulation in our previous study in the context of spatially-coupled edge caching provides another example~\cite{C_JGao2018}. Denote the number of all contents which are assigned with a nonnegative caching probability as $N_\mathrm{p}$. Then, the number of cache states to be considered decreases from $\binom{N_\mathrm{f}}{c}$ to $\binom{N_\mathrm{p}}{c}$, which is a significant reduction when $N_\mathrm{p}$ is much smaller than $N_\mathrm{f}$. Moreover, the ``very popular'' contents which are assigned with a caching probability of 1 also reduces the number of cache states to be considered. If $N_\mathrm{e}$ contents are assigned a caching probability in $(0, 1)$, then the number of cache states decreases to $\binom{N_\mathrm{e}}{c}$.

\emph{On the state level}: We have illustrated in Fig.~\ref{f:HeurisSolu} and the related example on how to choose a state caching probability vector that has less non-negative elements. On top of that, a further and more effective simplification can be used. Specifically, we could limit the considered states to a small number of states with large overall cached content request probabilities. This will significantly reduce the number of states to be considered and the resulting dimension of the state transition matrix. For example, if $N_\mathrm{e} = 100$ and $c = 20$, there are more than $10^{20}$ states. However, we can consider the top 1000 states that cache the most popular contents only. By setting a proper cut-off threshold, the proposed dynamic caching can still yield satisfactory performance. We will show this with an example in Section~\ref{s:Simu}, in which there are $10000$ contents but we only consider $30$ states.

%
%
%
%
%

\vspace{1mm}
\section{Numerical Examples}\label{s:Simu}

\begin{figure}[tt]
	\centering {\includegraphics[width=0.50\textwidth]{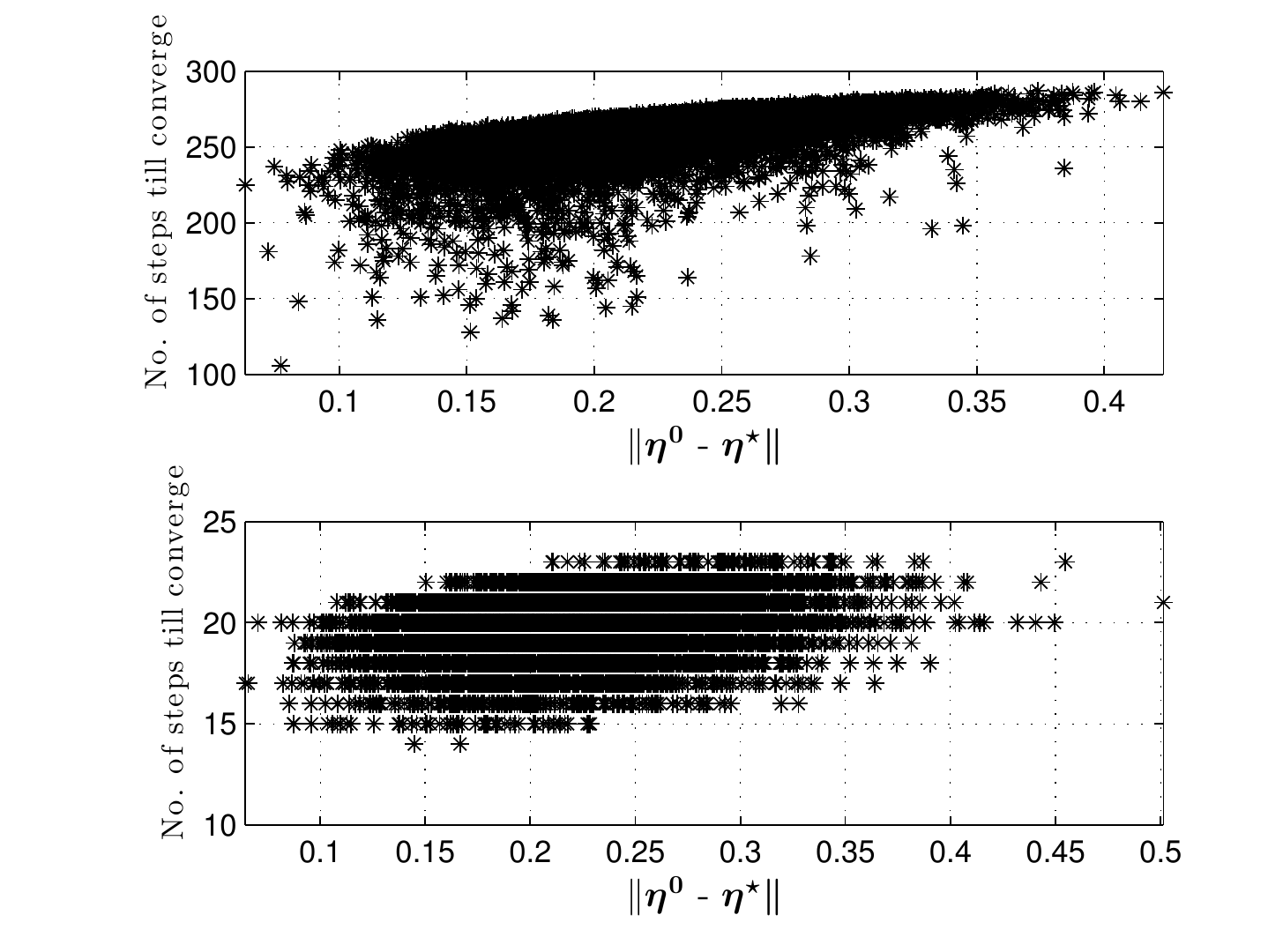}}
	\caption{Demonstration of the convergence speed of $\boldsymbol{\Theta}$ generated by Algorithms~\ref{a:Tgen}~and~\ref{a:RefineT} with 10000 random initial states.}\label{f:converge}
\end{figure}

\textit{Example 1: The convergence speed of the underlying Markov chain corresponding to the designed $\boldsymbol{\Theta}$}. In this illustrative example, 5 contents and a cache with size 2 is considered. Thus, there are 10 pure strategies and the mixed caching strategy is a probability vector with 10 elements. The transition probability matrix $\boldsymbol{\Theta}$ is first generated by Algorithms~\ref{a:Tgen} and then refined by Algorithms~\ref{a:RefineT}. For the purpose of illustrating the convergence performance of the proposed $\boldsymbol{\Theta}$, a constant instantaneous content popularity based on Zipf distribution is used in this example. The convergence speed of the underlying Markov chains of $\boldsymbol{\Theta}$ at the output of Algorithms~\ref{a:Tgen} and Algorithm~\ref{a:RefineT} are shown in the top and bottom subplots of Fig.~\ref{f:converge}, respectively. In each figure, 10000 tests with randomly generated initial $\boldsymbol{\eta}_0$ are conducted. 
Three observations can be made from Fig.~\ref{f:converge}. First, $\boldsymbol{\eta}$ always converge (in distribution) to the target $\boldsymbol{\eta}^\star$  with the designed replacement policy represented by $\boldsymbol{\Theta}$ in the 20000 tests regardless of the initial caching strategy, which validates the result in Theorem~\ref{t:PropTi}. Second, the convergence speed is shorter on average when the $\boldsymbol{\eta}^0$ is closer to $\boldsymbol{\eta}^\star$ and vice versa. Third, the refinement of $\boldsymbol{\Theta}$ by Algorithm~\ref{a:RefineT} significantly reduces the convergence speed, i.e., by a factor of 10.

\begin{figure}[tt]
	\centering {\includegraphics[width=0.45\textwidth]{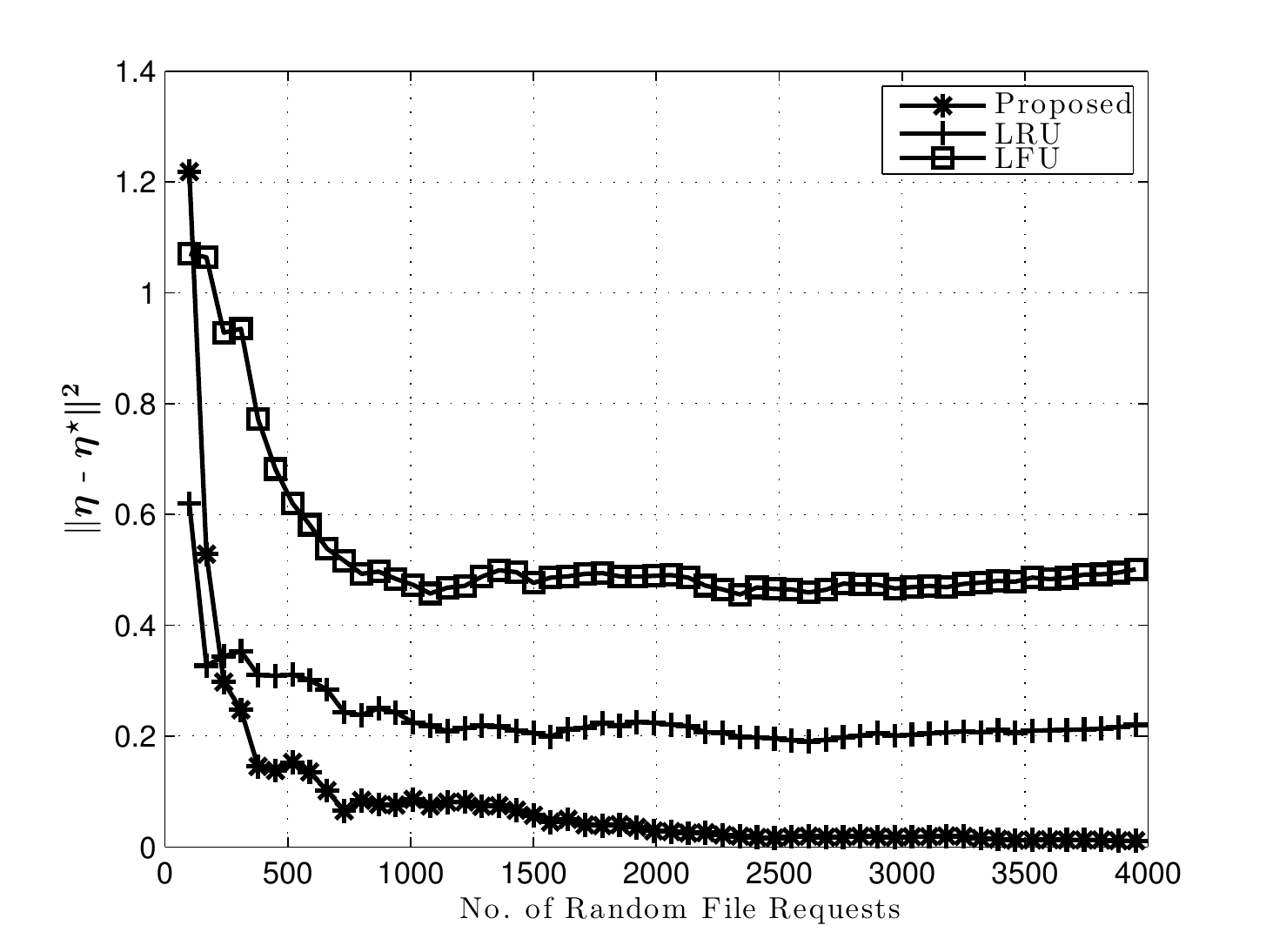}}
	\caption{Comparison of replacement policies: the proposed, LRU, and LFU.}\label{f:ReplacePolicy}
\end{figure}

\textit{Example 2: The comparison of the convergence (in distribution) of replacement policies}.  In this illustrative example, 15 contents and a cache with size 8 is considered. Thus, there are 6435 cache states, and the state caching probability vector has 6435 elements. The convergence performance of three replacement policies under constant content popularity is compared: the proposed policy corresponding to the designed $\boldsymbol{\Theta}$, LRU, and LFU. The content requests from UE are randomly generated and follows a Zipf distribution. The convergence is represented through the square norm of the difference between the current caching strategy $\boldsymbol{\eta}$ and the target caching strategy $\boldsymbol{\eta}^\star$ versus the number of content requests since the beginning of the simulations. The comparison of the convergence performance is shown in Fig.~\ref{f:ReplacePolicy}. It can be seen from this figure that the proposed replacement policy can converge to $\boldsymbol{\eta}^\star$ in distribution and thereby implement a given set of caching probabilities  when the instantaneous content popularity is a constant. By contrast, LRU or LFU cannot converge to $\boldsymbol{\eta}^\star$ under the same condition.

\begin{figure}[tt]
	\centering {\includegraphics[width=0.50\textwidth]{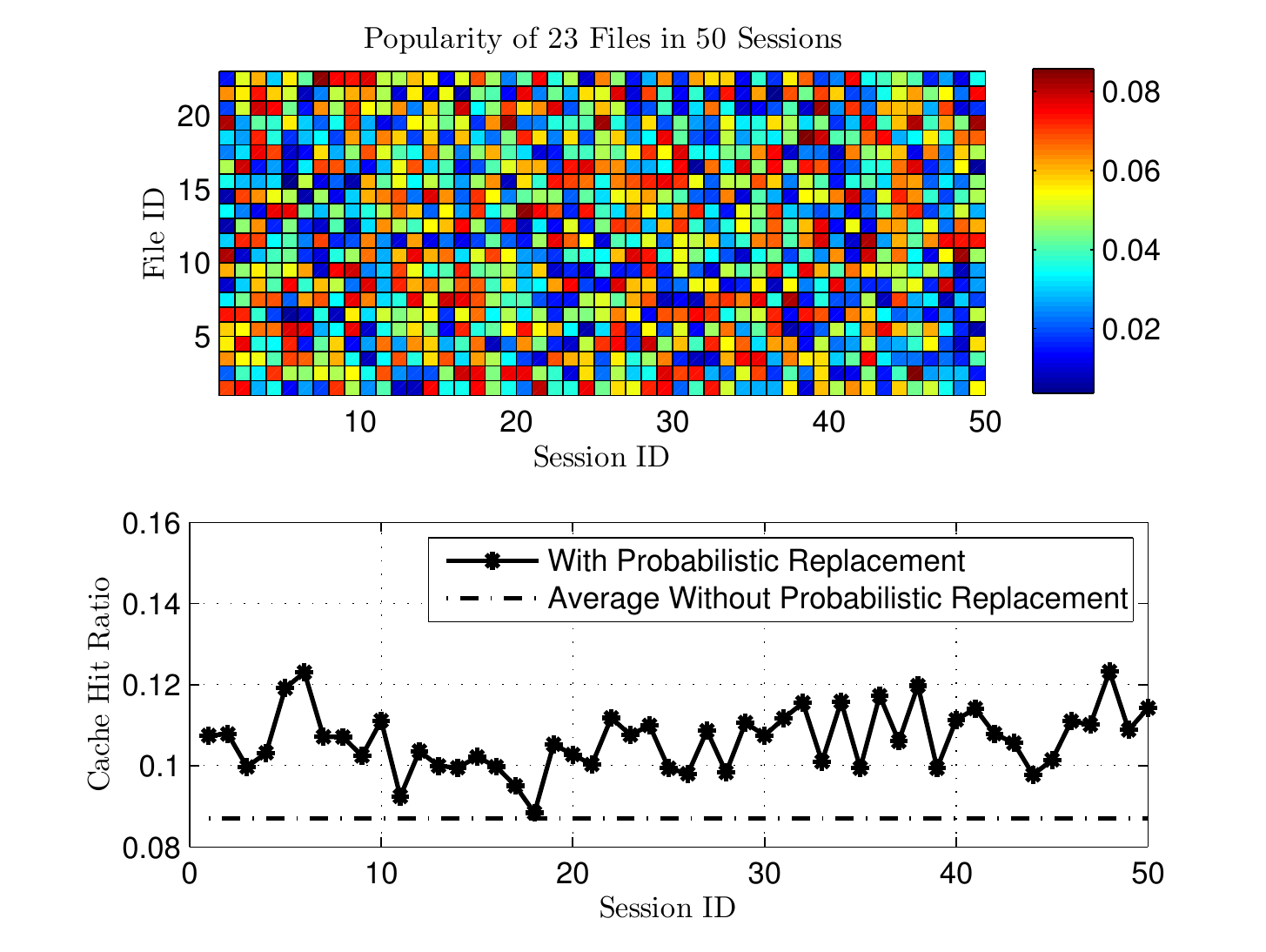}}
	\caption{Cache hit ratio with probabilistic replacement when content popularity varies over time - the case of random fluctuation.}\label{f:BenefitReplaceRandom}
\end{figure}

\begin{figure}[tt]
	\centering {\includegraphics[width=0.50\textwidth]{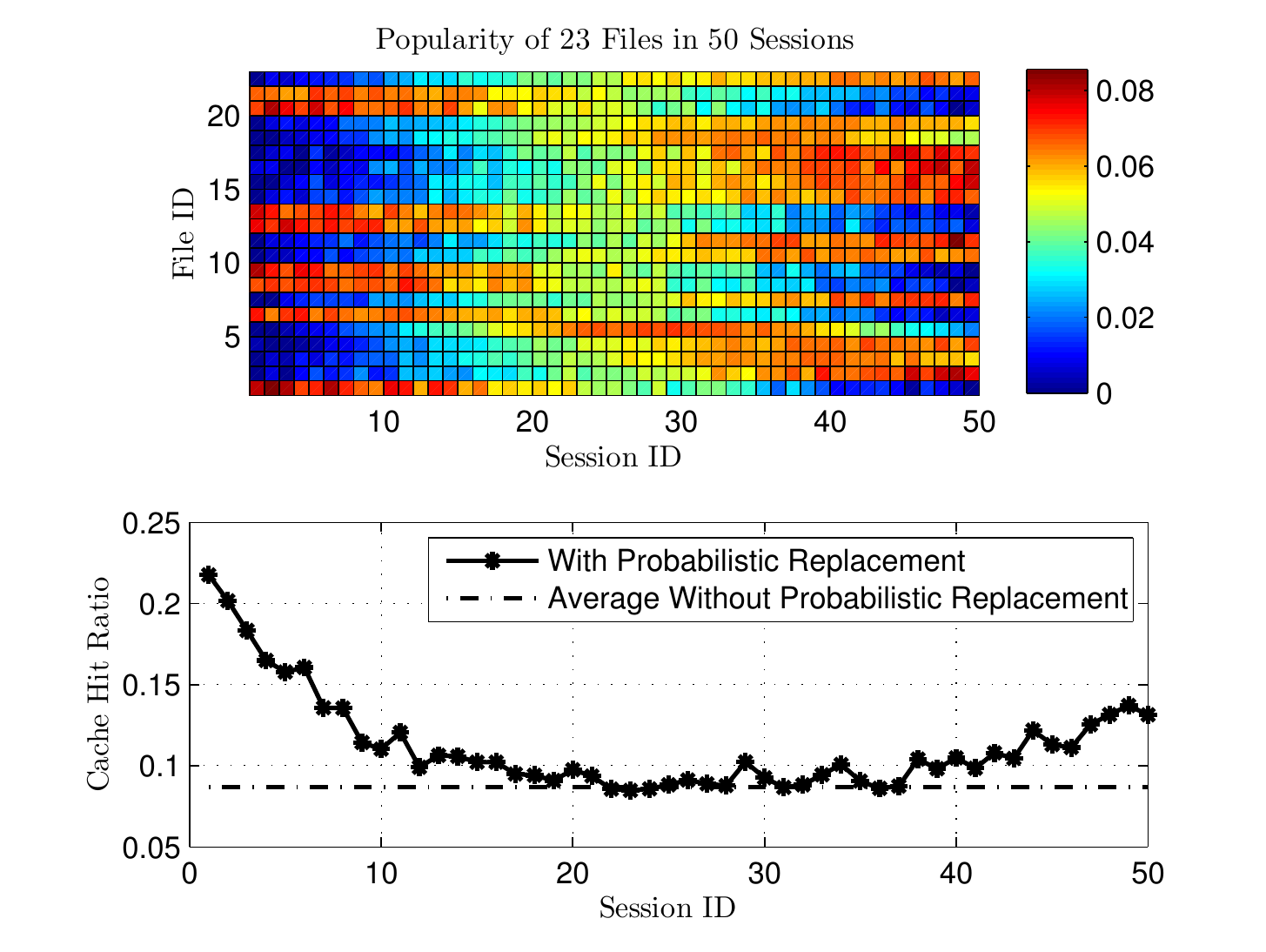}}
	\caption{Cache hit ratio with probabilistic replacement when content popularity varies over time - the case of smooth change.}\label{f:BenefitReplaceTrend}
\end{figure}

\textit{Example 3: The benefit of probabilistic content replacement}. In this illustrative example, the benefit of dynamic probabilistic content replacement with the designed replacement policy $\{\tau_{m, m^\prime}\}$ is demonstrated with 23 contents and a cache of size 2. A time duration divided into 50 sessions is considered, and $2\times 10^6$ content requests are generated in total. The 23 contents are equally popular overall but the popularity of each content varies in each session. Two different cases of variations in content popularity (in terms of content request probability) are considered: random fluctuation and smooth change, as shown in the top subplots of Figs.~\ref{f:BenefitReplaceRandom}~and~\ref{f:BenefitReplaceTrend}, respectively. The corresponding cache hit ratio by using dynamic probabilistic content replacement is given in the bottom subplots of Figs.~\ref{f:BenefitReplaceRandom}~and~\ref{f:BenefitReplaceTrend}, respectively. As the overall popularity is the same for each content, caching any two content without replacement would lead to a cache hit ratio of 2/23, or 0.087 approximately. It can be seen that from Figs.~\ref{f:BenefitReplaceRandom}~and~\ref{f:BenefitReplaceTrend} that the cache hit ratio is improved by using probabilistic content replacement in either case. The overall cache hit ratio is 0.1063 and 0.1085, equivalent to an increase of $22\%$ and $25\%$, in the cases of Figs.~\ref{f:BenefitReplaceRandom}~and~\ref{f:BenefitReplaceTrend}, respectively. The figures demonstrate that designed probabilistic content replacement based on the average content popularity can improve cache hit ratio by adapting to the varying content popularity when the instantaneous content popularity changes over time.

\textit{Example 4: The case of large content number and shot noise request model}. In this example, we consider a total of 10000 contents and demonstrate how to apply the proposed idea of dynamic caching when there are a large number of contents. The requests for the contents are generated in a window of 100 minutes based on the shot noise model using the exponential shape \cite{J_STraverso2013}. The average number of requests for each content in the considered time window follows a Zipf distribution with parameter $s$. The average content popularity is based on the overall number of requests in the considered 100 minutes. The first request for the contents follows a uniform distribution in $[0, 40]$ minutes. Contents have different lifetime, while the lifetime of every content is short so that almost all requests occur in the considered 100-minute time window. In such a scenario, there are two problems in applying the proposed dynamic caching. First, the number of cache state is prohibitively large due to the 10000 contents. Second, the optimal static caching strategy reduces to caching the $c$ most popular contents deterministically (i.e., $\eta_1^\star = 1$ and $\eta_l^\star = 0, \forall l > 1$). To apply and evaluate the idea of the proposed dynamic caching, we make the following two simplifications of the proposed design. First, we only consider 30 cache states with the largest static cache hit ratio. Second, in order to apply dynamic state transition, we use a non-optimal $\boldsymbol{\eta}^\star$ (of size $30\times 1$), in which the state caching probability of state $l$ is proportional to $\sum_{k \in \mathcal{G}^l} \varphi_k$. Note that the above two settings significantly simplify our original design and can lead to performance degradations. We compare the cache hit ratio of the optimal static caching, LRU, and the (simplified) proposed dynamic caching as well as the number of replacements of LRU and the (simplified) proposed dynamic caching under various cache sizes, values of $s$, and content lifetime (i.e., the duration between the first and the last requests).

\begin{figure}[!t]
	\centering \subfloat[Cache hit ratio versus popularity skewness parameter $s$.]
	{\includegraphics[angle=0,width=0.478\textwidth]{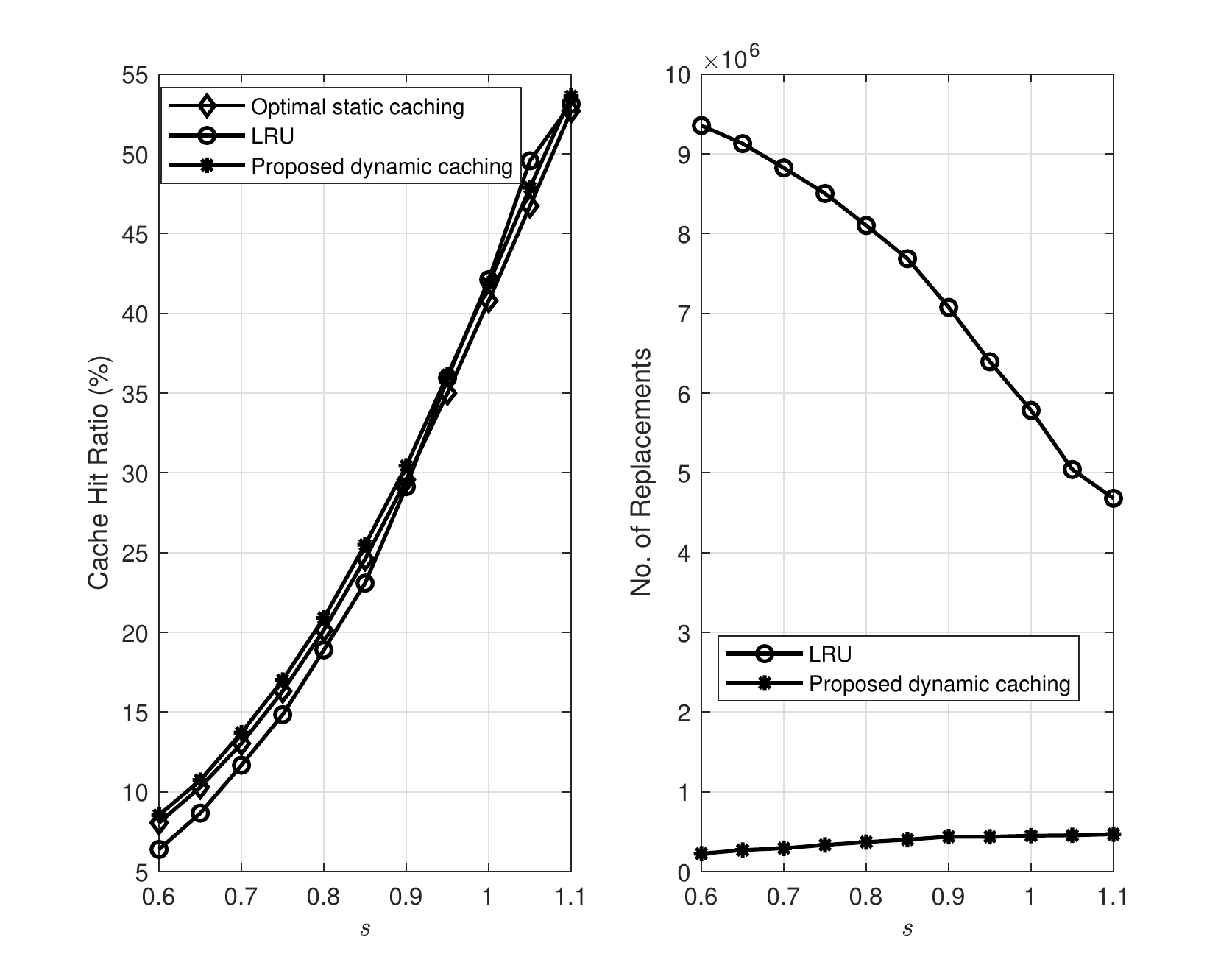}
		\label{f:SN_ZipfPara}}
	\hspace{1mm} 
	\subfloat[Cache hit ratio versus cache size $N_\mathrm{c}$.] 
	{\includegraphics[angle=0,width=0.478\textwidth]{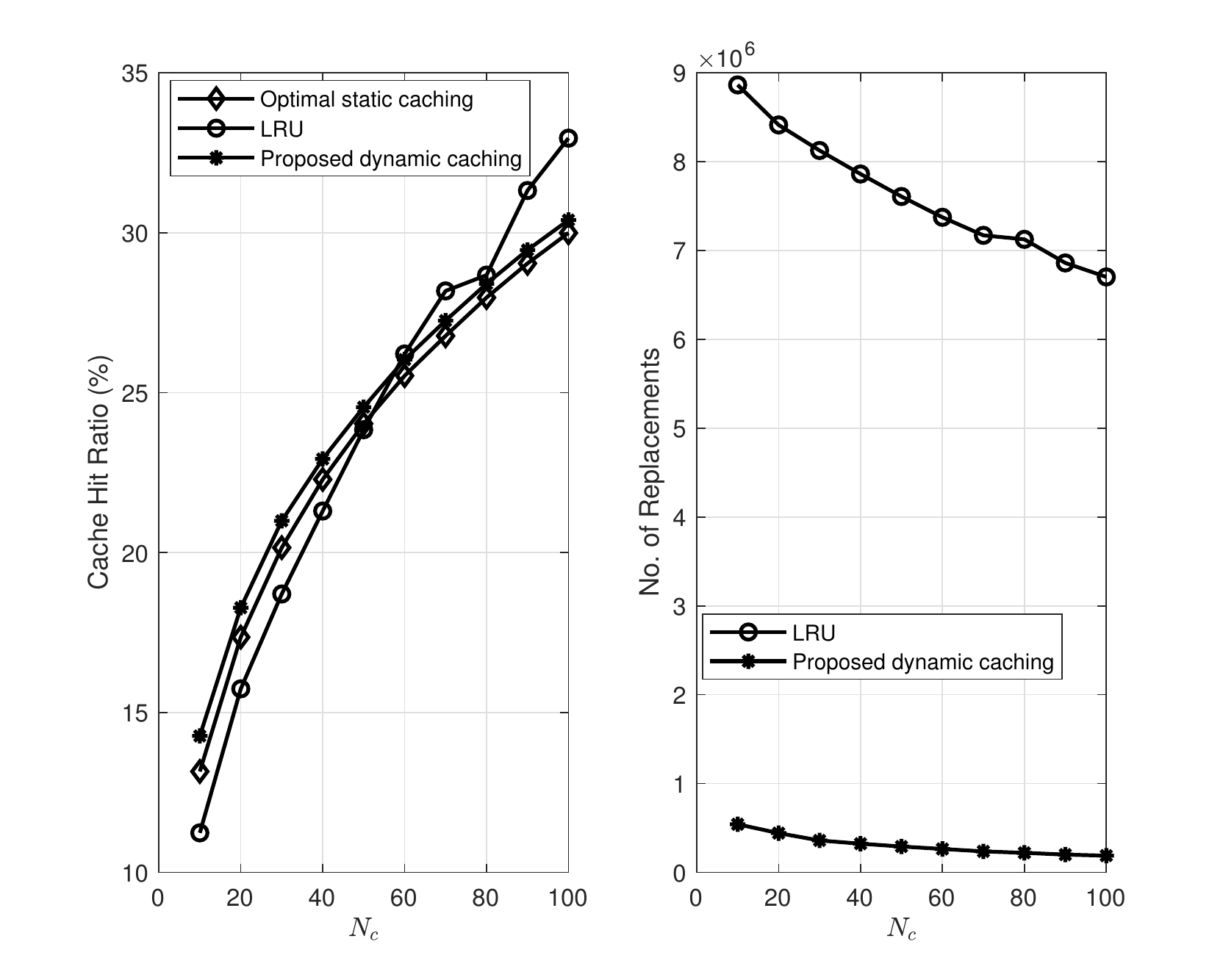}
		\label{f:SN_CacheSize}}
	\hspace{1mm} 
	\subfloat[Cache hit ratio versus popular content lifetime.]
	{\includegraphics[angle=0,width=0.478\textwidth]{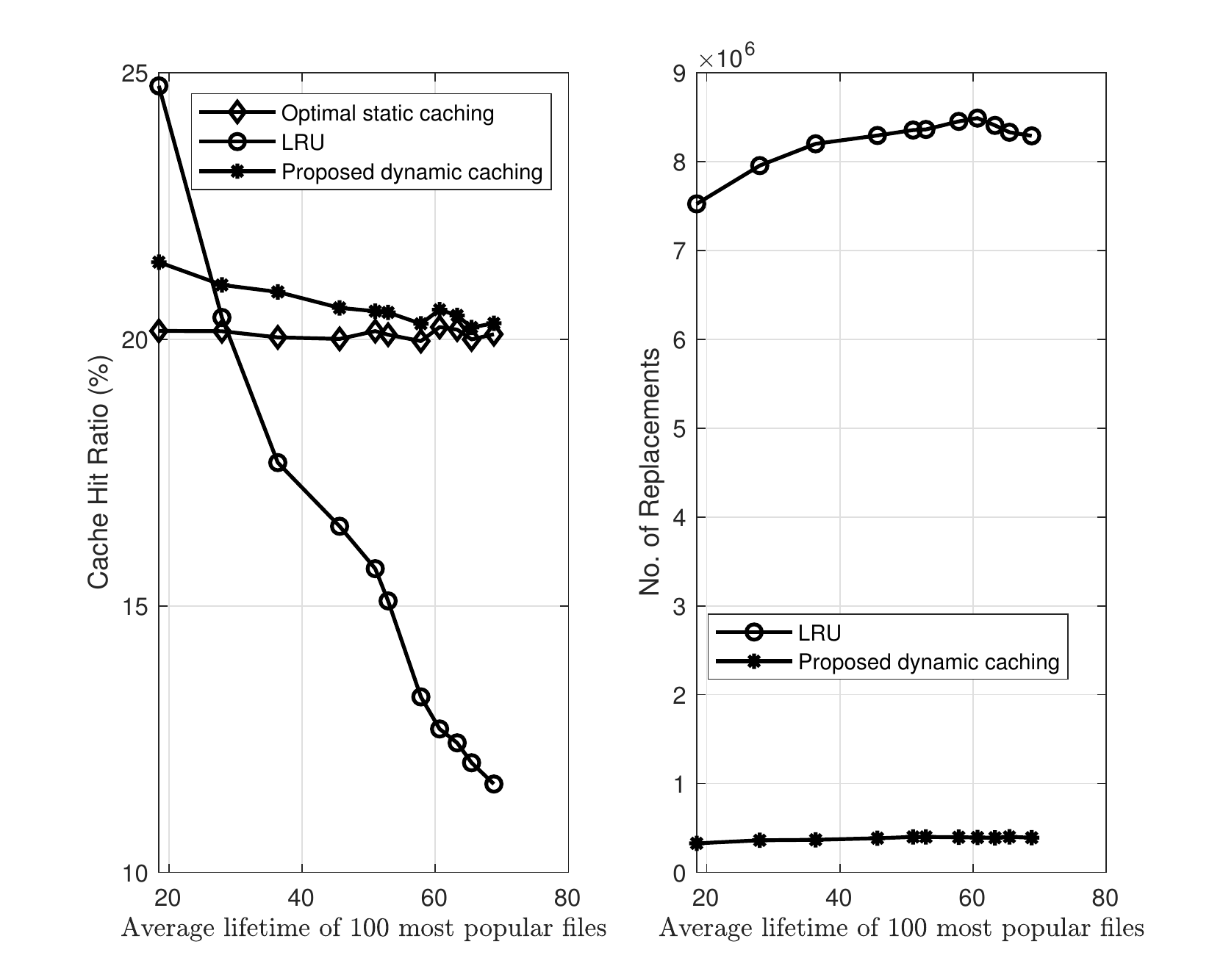}
		\label{f:SN_LifeTime}}
	\caption{Comparison of the optimal static caching, LRU, and the proposed dynamic caching under different settings.} 
\end{figure}

Fig.~\ref{f:SN_ZipfPara} shows the cache hit ratio of the three schemes as well as the number of replacements of LRU and the (simplified) proposed dynamic caching versus the content popularity skewness parameter $s$ in the range of $[0.6, 1.1]$. The cache size is 30, and the average content lifetime for the 100 most popular files is around $32.7$ minutes. Each point in the figure is an average over 240 runs of simulations, each of which randomly generates all content requests for the 10000 contents. The plot on the left-hand side shows that, when $s$ is small, the optimal static caching outperform LRU. Moreover, the proposed dynamic caching outperforms both LRU and the optimal static caching. The reason behind the two observation is that the replacement made by LRU can be less targeted when the skewness of popularity is small (i.e., the chance that the new content is less likely to be requested than the replaced content could be large) while the replacements made by the proposed method are guided by the average content popularity information. When $s$ becomes large and the number of ``most popular" contents become smaller, LRU can gain an advantage since all three schemes cache the most popular contents with large probabilities, but LRU has the strongest adaptivity and uses the rest of the cache to store temporarily popular contents. However, overall, the proposed dynamic caching, even after significant simplification, still outperforms both the optimal static caching and LRU in a wide range of $s$. Furthermore, the plot on the right-hand side shows that the proposed dynamic caching achieves such a performance with much less number of replacements compared to LRU (i.e., less than $1/40$ of the replacements by LRU when $s = 0.6$ and less than $1/8$ when $s= 1.1$). Therefore, the replacements in the proposed dynamic caching are more effective due to the exploitation of the average content popularity information.

Fig.~\ref{f:SN_CacheSize} shows the cache hit ratio of the three schemes as well as the number of replacements of LRU and the (simplified) proposed dynamic caching versus the cache size $N_\mathrm{c}$ in the range of $[10, 100]$. The popularity skewness parameter $s$ is 0.8, and the average content lifetime for the 100 most popular files is $32.7$ minutes. Each point in the figure is an average over 240 runs of simulations, each of which randomly generates all content requests for the 10000 contents. The logarithmic growth in the cache hit ratio versus the cache size shown in Fig.~\ref{f:SN_CacheSize} is consistent with the observations in existing research, e.g., \cite{C_AGhodsi2011}. The proposed dynamic caching always has an advantage over LRU in the cache hit ratio when the cache size $N_\mathrm{c}$ is no larger than 60 (i.e., 6\text{\textperthousand} of the size of all contents). In addition, the proposed dynamic caching always outperforms the optimal static caching. Furthermore, similar to the case shown in Fig.~\ref{f:SN_ZipfPara}, the proposed dynamic caching uses a significantly less number of replacements compared to that of LRU.

Fig.~\ref{f:SN_LifeTime} shows the cache hit ratio of the three schemes as well as the number of replacements of LRU and the (simplified) proposed dynamic caching versus the average content lifetime for the 100 most popular files. The popularity skewness parameter $s$ is 0.8, and the cache size $N_\mathrm{c}$ is 30. Each point in the figure is an average over 240 runs of simulations, each of which randomly generates all content requests for the 10000 contents. The left-hand side plot shows that, when the average lifetime of the 100 most popular contents is small, e.g., less than 30 minutes in the considered 100-minute window, dynamic caching (including LRU and the proposed caching) has an advantage over the static caching. This is because, given a fixed time window, a short lifetime can lead to a stronger temporal locality in content popularity. Moreover, when the content lifetime increases, the performance of LRU decreases very fast while the proposed dynamic caching still achieves a better cache hit ratio compared to the optimal static caching. The plot on the right-hand side shows again that the proposed dynamic caching uses a significantly less number of replacements than LRU.

\vspace{2mm}
\section{Conclusion}
In this work, we have studied dynamic probabilistic caching when the instantaneous content popularity may vary with time but the average content popularity is known. Specifically, we have designed probabilistic content placement and replacement policies with the objective of increasing cache hit ratio under varying instantaneous content popularity while converging to the target content caching probabilities under constant instantaneous content popularity. On the probabilistic content placement, we have established a novel connection between caching probabilities and probabilistic content placement policies through mixed strategies. On the probabilistic content replacement,  we have designed two algorithms to generate and refine the transition probability matrix so that the Markov chain has a unique steady state which achieves the target content placement. Our study has demonstrated that the proposed dynamic probabilistic caching can be applied on top of the existing results on finding the optimal caching probabilities to improve the cache hit ratio under time-varying content popularity while the optimal caching probabilities can be used as the target content caching probabilities. When combined with content popularity prediction, the proposed design can provide a competitive approach for dynamic caching under time-varying content popularity.

\setcounter{equation}{0}
\renewcommand{\theequation}{A.\arabic{equation}}

\section{Proof of Lemma~2}\label{s:ProofSProp}

For an arbitrarily given $n \times 1$ vector $\mathbf{x} = [x_1, \dots, x_{n}]^\mathrm{T}$ such that $x_k \geq 0, k \in \{1, \dots, n\}$ and $\|\mathbf{x}\| = 1$, it holds that:
\begin{align}\label{e:normSx}
\|\mathbf{S} \mathbf{x}\|^2 = \sum\limits_{l = 1}^{N_\mathrm{f}} (\mathbf{r}^l \mathbf{x})^2
\end{align}
where $\mathbf{r}^l$ is the $l$th row of matrix $\mathbf{S}$. It can be observed from $\mathbf{S}$ that each row $\mathbf{r}^l$ has $\binom{N_\mathrm{f}-1}{c-1}$ elements of the value 1 (and the rest of the elements are 0's). Moreover, it can also be seen that $\sum_{l = 1}^{N_\mathrm{f}} \mathbf{r}^l =  [c, c, \dots, c]$. Therefore, the norm in eq.~\eqref{e:normSx} can be rewritten as
\begin{align}\label{e:normSx2}
\|\mathbf{S} \mathbf{x}\|^2 &= \sum\limits_{l = 1}^{N_\mathrm{f}} \bigg( \sum\limits_{u = 1}^{\binom{N_\mathrm{f}-1}{c-1}}  x_{l,u}\bigg)^2 \nonumber \\
&\geq c (x_1^2 + \dots + x_n^2) = c
\end{align}
where $x_{l, u}$ is the element in $\mathbf{x}$ corresponding to the $u$th 1 in $\mathbf{r}^l$. The inequality in eq.~\eqref{e:normSx2} shows that for an arbitrarily given nonnegative and unit-size vector $\mathbf{x}$, the result after scaling and rotating $\mathbf{x}$ by the product $\mathbf{S} \mathbf{x}$ has a norm no less than $\sqrt{c}$. This is impossible if $\mathbf{S}$ has any singular value less than $\sqrt{c}$. \hfill$\blacksquare$ 

\setcounter{equation}{0}
\renewcommand{\theequation}{B.\arabic{equation}}

\section{Proof of Theorem~1}\label{s:ProofPropTi}

First, from the fact that $\boldsymbol{\Theta}$ is initialized to be $\mathbf{I}$ and the procedure~\eqref{e:BasicProcedure1}-\eqref{e:BasicProcedure6}, it can be seen that all elements of $\boldsymbol{\Theta}$ are in the range of $[0, 1]$ and that elements in the same column add up to 1 at any step. This proves that the generated $\boldsymbol{\Theta}$  is a stochastic matrix and thus a valid state transition matrix.

Second, it can be seen from the procedure~\eqref{e:BasicProcedure1}-\eqref{e:BasicProcedure6} that, with any $m$ and $m^\prime$ from $\{1, \dots, n\}$ and $m \neq m^\prime$, $\boldsymbol{\Theta}$ satisfies:
\begin{align}
\boldsymbol{\Theta}(m, m^\prime) \leq \varphi_{ m, m^\prime}.
\end{align} 
As a result, 
\begin{align}
\tau_{m, m^\prime} = \frac{\boldsymbol{\Theta}(m, m^\prime)}{\varphi_{ m, m^\prime}} \leq 1.
\end{align}

Third, it is obvious that $\boldsymbol{\Theta} = \mathbf{I}$ satisfies eq.~\eqref{e:TransSteady}.  The update of elements in~\eqref{e:BasicProcedure2}-\eqref{e:BasicProcedure6} satisfies:
\begin{align}
\Delta(\boldsymbol{\Theta}(m^\prime, m)) \eta^m + \Delta(\boldsymbol{\Theta}(m^\prime, m^\prime)) \eta^{m^\prime} = 0,
\end{align} 
where $\Delta(\boldsymbol{\Theta}(m^\prime, m))$ is the change of $\boldsymbol{\Theta}(m^\prime, m)$ (i.e., from 0 to $\delta$), and $\Delta(\boldsymbol{\Theta}(m^\prime, m^\prime))$ is the change of $\boldsymbol{\Theta}(m^\prime, m^\prime)$. Accordingly, the update of elements in the $m^\prime$th row does not break the corresponding $m^\prime$th equality in eq.~\eqref{e:TransSteady}. Similarly, it holds that: 
\begin{align}
\Delta(\boldsymbol{\Theta}(m, m^\prime)) \eta^{m^\prime} + \Delta(\boldsymbol{\Theta}(m, m)) \eta^{m} = 0.
\end{align} 
Thus, the updates of elements in the $m$th row does not break the corresponding $m$th equality in eq.~\eqref{e:TransSteady}. It follows that the procedure in~\eqref{e:BasicProcedure2}-\eqref{e:BasicProcedure6} sustains the equality in eq.~\eqref{e:TransSteady}. Therefore, eq.~\eqref{e:TransSteady} is satisfied at any step.

Last, it can be seen that the underlying Markov chain constructed while generating the state transition matrix $\boldsymbol{\Theta}$ is irreducible and ergodic. Therefore, there is a unique steady state vector. Given eq.~\eqref{e:TransSteady}, it can be seen that $\boldsymbol{\eta}^\star$ is the unique steady state vector.  \hfill$\blacksquare$
\vspace{0mm}

\setcounter{equation}{0}
\renewcommand{\theequation}{C.\arabic{equation}} 

\section{Proof of Lemma~3}

The instantaneous state transition matrix at the $q$th request is: 
\begin{align}
\boldsymbol{\Theta}^{(q)} = \sum\limits_{k = 1}^{N_\mathrm{f}} \varphi_k^{(q)} \boldsymbol{\Theta}_k,
\end{align}
where $\boldsymbol{\Theta}_k$ is independent on the instantaneous content popularity. The overall state transition matrix averaged over the $N_\mathrm{r}$ requests is therefore:
\begin{align}
\bar{\boldsymbol{\Theta}} = \frac{1}{N_\mathrm{r}} \sum\limits_{q =1}^{N_\mathrm{r}} \sum\limits_{k = 1}^{N_\mathrm{f}} \varphi_k^{(q)} \boldsymbol{\Theta}_k = \sum\limits_{k = 1}^{N_\mathrm{f}} \left(\frac{1}{N_\mathrm{r}} \sum\limits_{q =1}^{N_\mathrm{r}} \varphi_k^{(q)}\right)  \boldsymbol{\Theta}_k.
\end{align}
The item in the brackets represent the average content request probabilities. Therefore, it follows that:
\begin{align}
\bar{\boldsymbol{\Theta}} = \sum\limits_{k = 1}^{N_\mathrm{f}} \varphi_k \boldsymbol{\Theta}_k = \boldsymbol{\Theta}. 
\end{align}
This completes the proof. \hfill$\blacksquare$

\balance


\begin{IEEEbiography}[{\includegraphics[width=1in,height=1.25in,clip,keepaspectratio]{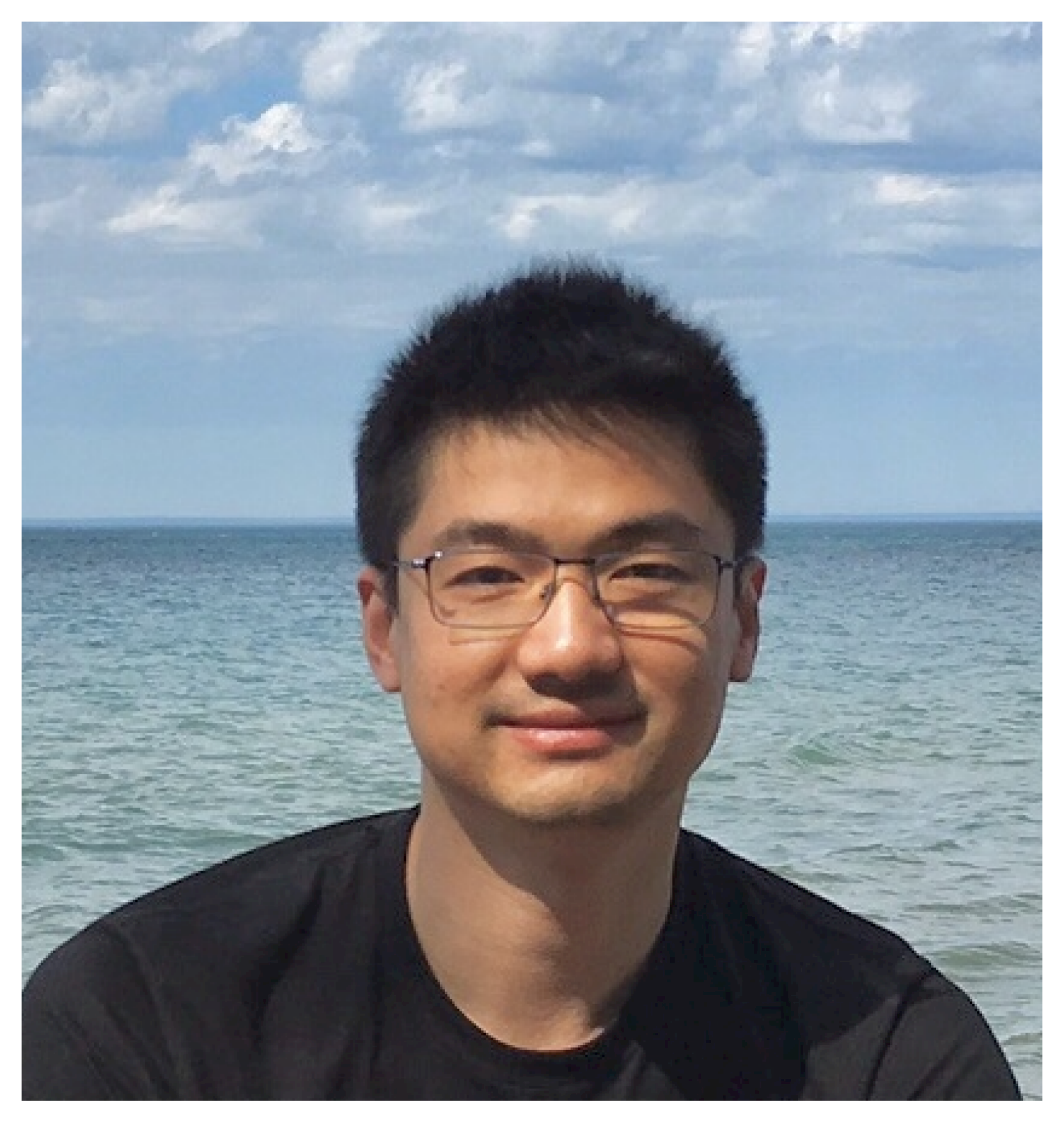}}]{Jie Gao}
		(S'09-13, M'17) received the B.Eng. degree in Electronics and Information Engineering from Huazhong University of Science and Technology, Wuhan, China, in 2007, and the M.Sc. and Ph.D. degrees in Electrical Engineering from the University of Alberta, Edmonton, Alberta, Canada, in 2009 and 2014, respectively. He is currently a research associate at the Department of Electrical \& Computer Engineering at the University of Waterloo. Dr.~Gao was a recipient of Natural Science and Engineering Research Council of Canada Postdoctoral Fellowship, Ontario Centres of Excellence TalentEdge Fellowship, and Alberta Innovates-Technology futures graduate student scholarship. His research interests include the application of optimization, game theory and mechanism design, and machine learning in wireless communication networks and systems.
\end{IEEEbiography}


\begin{IEEEbiography}[{\includegraphics[width=1in,height=1.25in,clip,keepaspectratio]{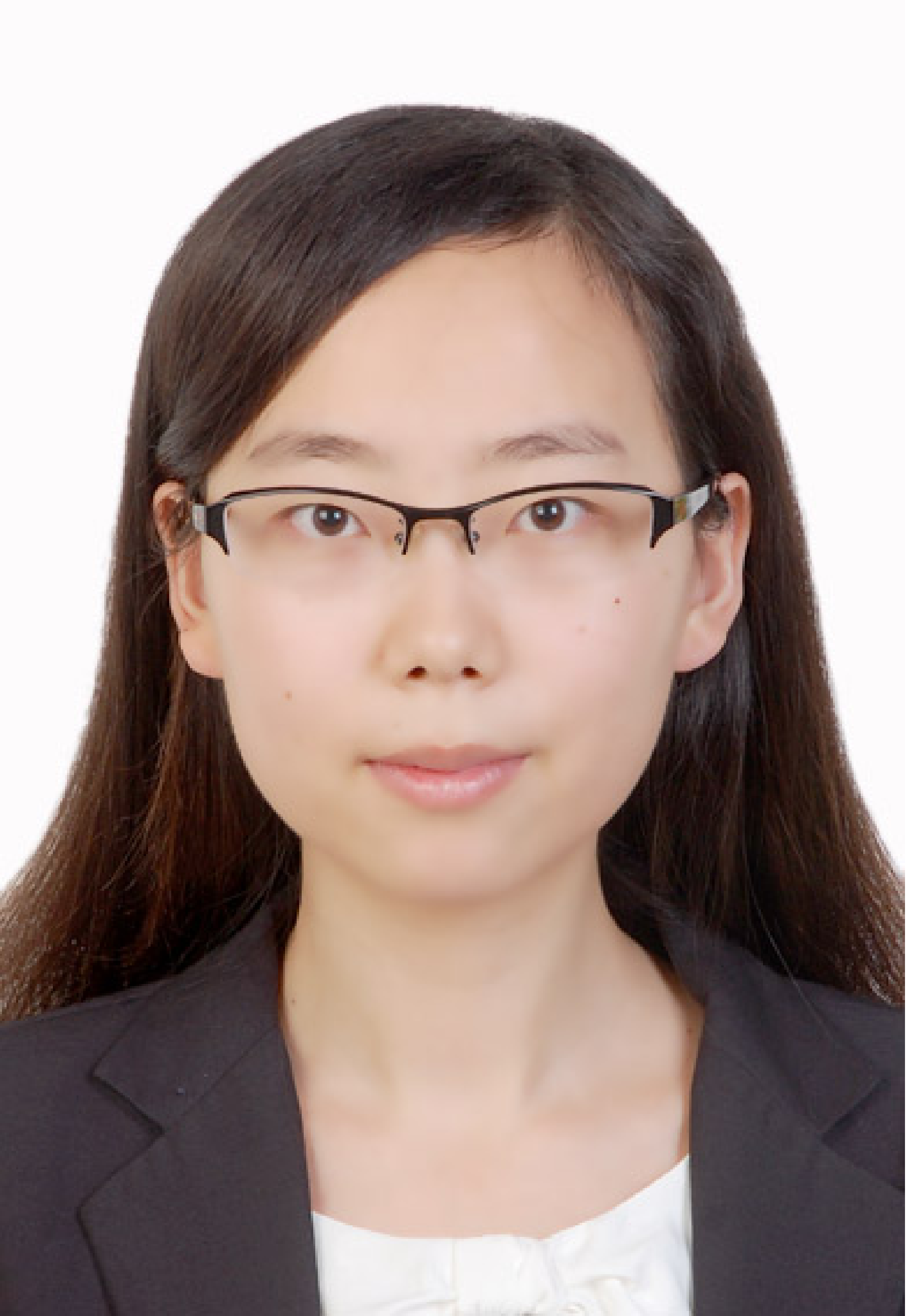}}]{Shan Zhang} 
	(S'13-M'16) received Ph.D. degree in electronic engineering from Tsinghua University, Beijing, China, in 2016. She is currently an assistant professor in the School of Computer Science and Engineering, Beihang University, Beijing, China. She was a post doctoral fellow in Department of Electronical and Computer Engineering, University of Waterloo, Ontario, Canada, from 2016 to 2017. Her research interests include mobile edge computing, wireless network virtualization and intelligent management. She received the Best Paper Award at the Asia-Pacific Conference on Communication in 2013.
\end{IEEEbiography}

\vspace{2cm}
\vfill

\begin{IEEEbiography}[{\includegraphics[width=1in,height=1.25in,keepaspectratio]{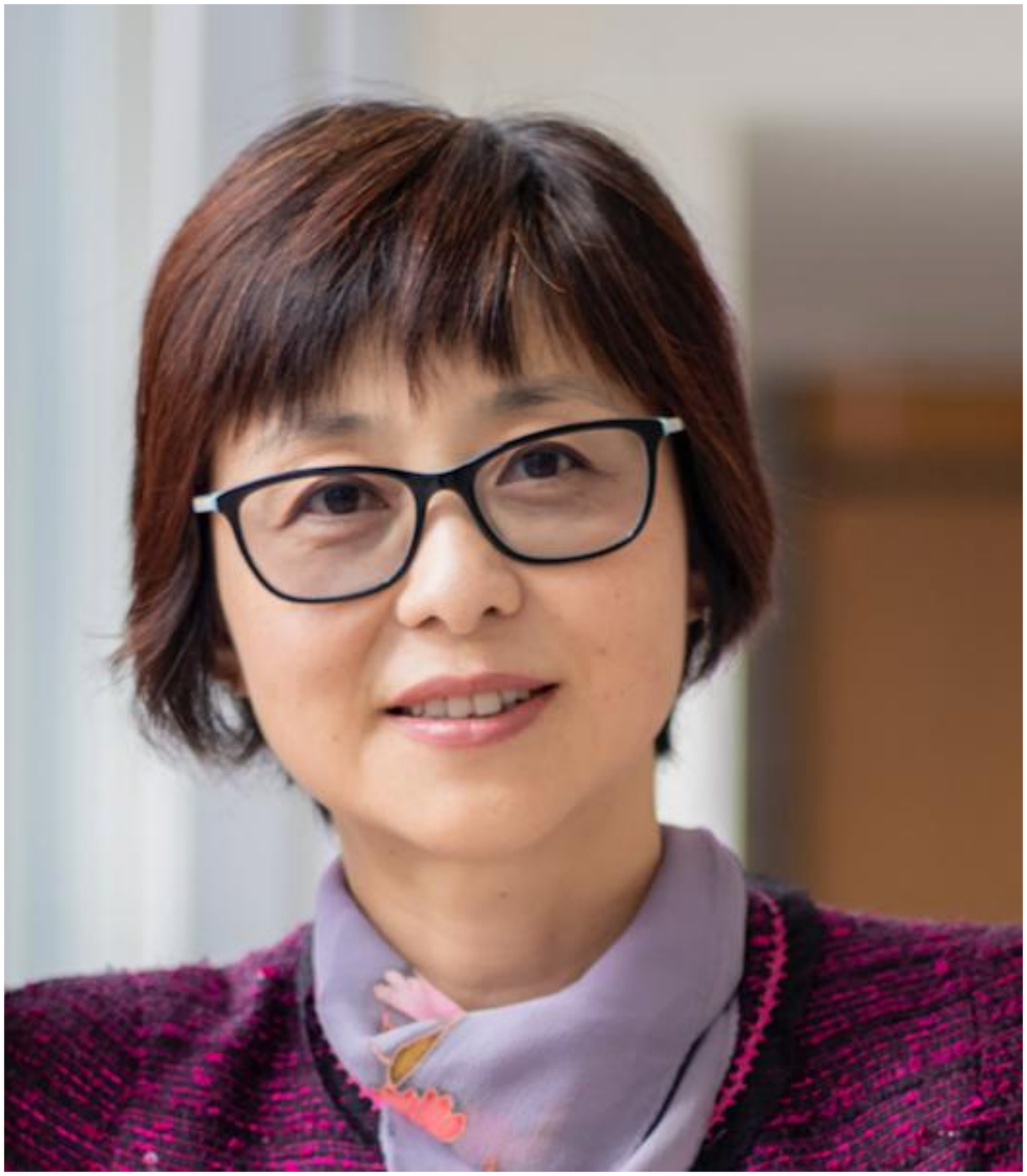}}]{Lian Zhao}
(S99-M03-SM06) received the Ph.D. degree from the Department of Electrical and Computer Engineering (ELCE), University of Waterloo, Canada, in 2002. She joined the Department of Electrical and Computer Engineering at Ryerson University, Toronto, Canada, in 2003 and has been a Professor in 2014. Her research interests are in the areas of wireless communications, radio resource management, mobile edge computing, caching and communications, and vehicular ad-hoc networks. \\
She has been selected as an IEEE Communication Society (ComSoc) Distinguished Lecturer (DL) for 2020 and 2021; received the Best Land Transportation Paper Award from IEEE Vehicular Technology Society in 2016; Top 15 Editor in 2015 for IEEE Transaction on Vehicular Technology; Best Paper Award from the 2013 International Conference on Wireless Communications and Signal Processing (WCSP) and Best Student Paper Award (with her student) from Chinacom in 2011; the Canada Foundation for Innovation (CFI) New Opportunity Research Award in 2005. She has been serving as an Editor for IEEE Transactions on Vehicular Technology, IEEE Transactions on Wireless Communications, IEEE Internet of Things Journal. She served as co-General Chair for IEEE GreeenCom 2018; co-Chair for IEEE Globecom 2020 and IEEE ICC 2018 Wireless Communication Symposium; workshop co-Chair for IEEE/CIC ICCC 2015; local arrangement co-Chair for IEEE VTC Fall 2017 and IEEE Infocom 2014; co-Chair for IEEE Global Communications Conference (GLOBECOM) 2013 Communication Theory Symposium. She served as a committee member for NSERC (Natural Science and Engineering Research Council of Canada) Discovery Grants Evaluation Group for Electrical and Computer Engineering 2015 to 2018; a Chair for the Professional Relations Committee (PRC) for IEEE Canada for 2020. She is a licensed Professional Engineer in the Province of Ontario, a senior member of the IEEE Communication and Vehicular Society. 
\end{IEEEbiography}

\vspace{-3cm}

\begin{IEEEbiography}[{\includegraphics[width=1in,height=1.3in,clip,keepaspectratio]{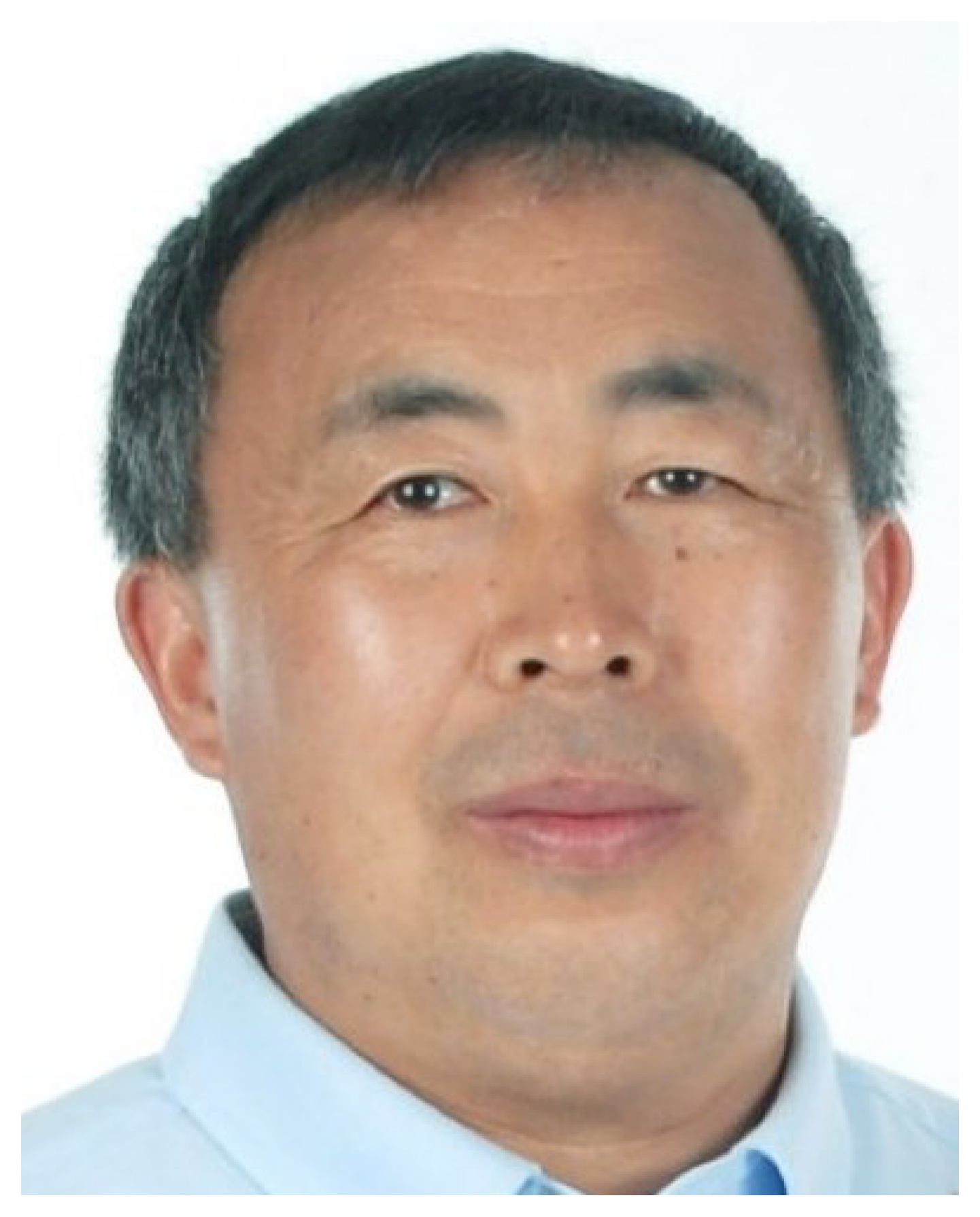}}]{Xuemin (Sherman) Shen} (IEEE M’97-SM’02- F09) received the  B.Sc.(1982) degree from Dalian Maritime University (China) and the M.Sc. (1987) and Ph.D. degrees (1990) from Rutgers University, New Jersey (USA), all in electrical engineering.He is a University Professor and the Associate Chair for Graduate Studies, Department of Electrical and Computer Engineering, University of Waterloo, Canada. Dr. Shen’s research focuses on resource management, wireless network security, social networks, smart grid, and vehicular ad hoc and sensor networks. Dr. Shen served as the Technical Program Committee Chair/CoChair for IEEE Globecom’16, Infocom’14, IEEE VTC’10 Fall, and Globecom’07, the Symposia Chair for IEEE ICC’10, the Tutorial Chair for IEEE VTC’11 Spring and IEEE ICC’08, the General Co-Chair for ACM Mobihoc’15, Chinacom’07 and the Chair for IEEE Communications Society Technical Committee on Wireless Communications. He also serves/served as the Editor-in-Chief for IEEE Internet of Things Journal, IEEE Network, Peerto-Peer Networking and Application, and IET Communications; a Founding Area Editor for IEEE Transactions on Wireless Communications; an Associate Editor for IEEE Transactions on Vehicular Technology, Computer Networks, and ACM/Wireless Networks, etc.; and the Guest Editor for IEEE JSAC, IEEE Wireless Communications, and IEEE Communications Magazine, etc. Dr. Shen received the Excellent Graduate Supervision Award in 2006, and the Premiers Research Excellence Award (PREA) in 2003 from the Province of Ontario, Canada. He is a registered Professional Engineer of Ontario, Canada, an Engineering Institute of Canada Fellow, a Canadian Academy of Engineering Fellow, a Royal Society of Canada Fellow, and a Distinguished Lecturer of IEEE Vehicular Technology Society and Communications Society.
\end{IEEEbiography}

\end{document}